\documentclass[12pt]{article}
\usepackage{fullpage}
\usepackage[hyphens]{url}
\usepackage{hyperref}
\usepackage[T1]{fontenc}    
\usepackage{booktabs}       
\usepackage{amsfonts}       
\usepackage{nicefrac}       
\usepackage{microtype}      
\usepackage{tcolorbox}
\usepackage{diagbox}
\usepackage{enumerate}
\usepackage[shortlabels]{enumitem}
\usepackage{algorithmic}
\usepackage{algorithm}
\usepackage{subcaption}
\usepackage{tabularx}
\usepackage{verbatim}
\usepackage{natbib}
\usepackage{tikz}
\usepackage{tikz-cd}
\usepackage{placeins} 

\usepackage{graphicx} 
\usepackage{caption}
\usepackage{amsmath}
\usepackage{amsthm}
\usepackage{amssymb}
\usepackage{tablefootnote}
\usepackage{multirow}
\usepackage{enumerate}
\usepackage{color}
\usepackage{xcolor}

\allowdisplaybreaks[4]

\usepackage{mathrsfs}

\usepackage{bm,todonotes}

\allowdisplaybreaks

\usepackage{amsthm}

\usepackage{xr}
\newtheorem{rem}{Remark}[section]
\newtheorem{ass}{Assumption}
\newtheorem{theorem}{Theorem}

\newtheorem{corollary}{Corollary}
\newtheorem{proposition}{Proposition}

\newtheorem{lemma}{Lemma}
{
	\theoremstyle{definition}
	\newtheorem{definition}{Definition}
	\newtheorem{example}{Example}

}

\definecolor{wjs}{RGB}{200,0,50}

\definecolor{hyw}{RGB}{153,000,000}

\long\def\new#1{\bgroup\color{black}#1\egroup}

\hypersetup{
	colorlinks=true,
	filecolor=black,
	citecolor=blue,
	urlcolor=black,
}

\newcommand{\beq}{\begin{equation}}
	\newcommand{\eeq}{\end{equation}}
\newcommand{\beas}{\begin{align*}}
	\newcommand{\eeas}{\end{align*}}
\newcommand{\bea}{\begin{align}}
	\newcommand{\eea}{\end{align}}
\newcommand{\bei}{\begin{itemize}}
	\newcommand{\eei}{\end{itemize}}
\newcommand{\ben}{\begin{enumerate}}
	\newcommand{\een}{\end{enumerate}}
\newcommand{\bet}{\begin{theorem}}
	\newcommand{\eet}{\end{theorem}}
\newcommand{\bel}{\begin{lemma}}
	\newcommand{\eel}{\end{lemma}}
\newcommand{\bep}{\begin{proposition}}
	\newcommand{\eep}{\end{proposition}}
\newcommand{\bed}{\begin{definition}}
	\newcommand{\eed}{\end{definition}}
\newcommand{\bec}{\begin{corollary}}
	\newcommand{\eec}{\end{corollary}}
\newcommand{\bex}{\begin{example}}
	\newcommand{\eex}{\end{example}}

\newcommand{\Expect}{\mathbb{E}}
\newcommand{\Cov}{\mathrm{Cov}}

\numberwithin{equation}{section}

\def\bSig\mathbf{\Sigma}

\newcommand{\wh}{\widehat} 
\newcommand{\wt}{\widetilde} 

\newcommand{\bbR}{\mathbb{R}}

\newcommand{\sP}{\mathcal{P}}

\newcommand{\vh}{\mathbf{h}}

\newcommand{\vs}{\mathbf{s}}

\newcommand{\vmu}{\boldsymbol{\mu}}

\begin{document}
	
\title{Differential Density Analysis in Single-Cell Genomics Using Specially Designed Exponential Families}

\author{Hanxuan Ye$^*$, \ Zachary Qian$^*$ \ and \ Hongzhe Li}
\date{}
\maketitle
\thispagestyle{empty}
\footnotetext{Hanxuan Ye is Postdoctoral Fellow, Department of Biostatistics, Epidemiology and Informatics, Perelman School of Medicine, University of Pennsylvania, PA 19104 (E-mail: \emph{Hanxuan.Ye@Pennmedicine.upenn.edu}). Zachary Qian is PhD Candidate, Department of Biostatistics, Epidemiology and Informatics, Perelman School of Medicine, University of Pennsylvania, Philadelphia, PA 19104 (E-mail: \emph{Zachary.Qian@Pennmedicine.upenn.edu}).  Hongzhe Li is Professor of Biostatistics and Statistics, Department of Biostatistics, Epidemiology and Informatics, Perelman School of Medicine, University of Pennsylvania, Philadelphia, PA 19104 (E-mail: \emph{hongzhe@upenn.edu}). $^*$: Ye and Qian contributed equally to this work. }

\begin{abstract}
	Recent advances in high-resolution sequencing have paved the way for population-scale analysis in single-cell RNA-sequencing (scRNA-seq) data. scRNA-seq data, in particular, have proven to be extremely powerful in profiling a variety of outcomes such as disease and aging. The abundance of scRNA-seq data makes it possible to model each individual's gene expression as a probability density across cells, offering a richer representation than summary statistics such as means or variances, and allowing for more nuanced group comparisons. To this end, we propose a model-agnostic framework for density estimation and inference based on specially designed exponential families~(SEF), which accommodates diverse underlying models without requiring prior specifications. The proposed method enables estimation and visualization for both individual-specific and group-level gene expression densities, as well as conducting formal hypothesis testing for expression density difference across groups of interest. It relies on relaxed assumptions with established asymptotic properties and a consistent covariance estimator for valid inference. Through simulation under various scenarios, the SEF-based approach demonstrates good error control and improved statistical power over competing methods,including pseudo-bulk tests and moment estimators. Application to a population-scale scRNA-seq dataset from patients with systemic lupus erythematosus  identified genes and gene sets that are missed from pseudo-bulk based tests.  
\end{abstract}


\section{Introduction}
\label{s:intro}

Recent advances in high-resolution sequencing have enabled population-scale analysis of single-cell RNA sequencing (scRNA-seq) data. These datasets profile gene expression at single-cell resolution across numerous individuals, offering unprecedented opportunities to study complex biological processes such as disease progression and aging. 
A key scientific objective in such studies is to detect differential expression~(DE) across groups (e.g., case vs. control). Existing approaches to single-cell DE (scDE) analysis focus exclusively on mean expression and can be dichotomized into two categories: \textit{pseudo-bulk} and \textit{cell-level methods}. 
Pseudo-bulk methods aggregate cell-level counts to the sample level and apply bulk RNA-seq tools such as DESeq2~\citep{love2014} and edgeR~\citep{robinson2010edger}, or nonparametric tests such as Wilcoxon rank-sum, t-test, or Kolmogorov-Smirnov (KS) goodness-of-fit. While computationally efficient, these approaches reduce rich single-cell information to a single summary statistic, obscuring variation in shape, tails, and modality. This makes them fail to capture more subtle distributional changes beyond the basic statistics.
Cell-level methods retain cell-level resolution, either via parametric models, e.g. MAST~\citep{finak2015} and Monocle~\citep{trapnell2014Monocle1, qiu2017monocle2}, which fits a hurdle model to log-transformed normalized counts, or via nonparametric comparisons to account for differences in modalities~\citep{wang2018sigemd}, including optimal transport–based methods such as waddR and BSDE~\citep{schefzik2021waddr, zhangm2022} opt to use a component mixture model as well to handle dropout and technical noise. However, parametric methods risk model misspecification, and many non-parametric approaches remain computationally costly at realistic population scale and lack tractable asymptotics for inference.

Besides DE detection of mean expression difference, the abundance of single-cell data naturally motivates estimation of the entire distribution  of gene expression across cells for a given cell type. 
There is a growing evidence of transcriptome variation across individual cells of the same cell type \citep{AltschulerWu2015,Upadhya2025,Liu2023,Gatlin2025}. While such  expression variation may be a byproduct of  dynamic or homeostatic processes,  single‐cell molecular variation  be crucial for population‐level function \citep{AltschulerWu2015}. For example, \cite{Upadhya2025} showed that human neurodevelopmental conditions can drive increased gene expression variability in brain cell types, with the potential to contribute to diverse phenotypic outcomes.
Besides mean and variance, 
both individual- and group-level probability distribution  estimates can reveal subtle expression changes, characterize heterogeneity, and provide a powerful visualization tool. Yet, few existing methods directly estimate these probability distributions  while also enabling valid statistical inference for group differences.

In this work, we propose a novel framework that treats the distribution   of expression counts across cells within an individual for a given gene as a random probability distribution. We estimate  the cell-type specific gene expression density using a flexible semi-parametric method inspired by the specially designed exponential family~(SEF) density estimation of~\cite{efron96}. The technique builds upon kernel density estimation~(KDE) but augments it with an exponential tilting term that enables more efficient estimation of individual- and group-level densities. This hybrid approach inherits the flexibility of nonparametric KDE while achieving improved statistical accuracy. Pure parametric models may fail to capture the complexity of real scRNA-seq distributions, especially when the parametric family is misspecified.
On the other hand, nonparametric KDE methods, though flexible, incur bias in settings with moderate cell counts. The SEF approach provides a compelling middle ground. It achieves lower bias and better approximation of true densities compared to KDE, as demonstrated in prior work~\citep{hjort1995nonparametric, hjort1996performance}. We anticipate that these properties make it particularly well-suited for single-cell applications.

Concretely, we extend Efron's SEF framework to the multigroup setting by modeling the expression counts of all cells within each individual as a random distribution, and estimating both individual-specific and group-level mean densities. We first aggregate expression data across the cells of all individuals to estimate a shared baseline~(the carrier density) using KDE, and then model individual and group mean densities as an exponential tilt of the carrier~(after normalization). Parameters are estimated via likelihood maximization, equivalent to moment matching under the SEF structure.  The estimated coefficients for different densities, within this model framework, carry interpretable meaning: they are identical given the same expression distribution. Meanwhile, with rich enough sufficient statistics, they differ as long as the distributions under comparison are distinct. Thus, the SEF test, a test based on the SEF estimator can distinguish the distributions as long as they differ in any moment.   We establish the asymptotic properties of the SEF estimator, including consistency and asymptotic normality. A consistent estimator for its covariance is also derived, enabling rigorous hypothesis testing for differential expression distribution across groups based on the estimated coefficients.

Compared to existing parametric models such as  Poisson-Gamma and  Zero-Inflated Negative Binomial~(ZINB) models that have been  applied  for single-cell data, our proposed  methods mitigates the risk of model misspecification, and also allows flexible choice of sufficient statistics to target moments of interest.  In addition,  the estimation procedure is easy to implement, as most computation can be done using a Poisson regression model with binned data, originally introduced by~\cite{lindsey1974comparison,lindsey1974construction}. The choice of kernel can be tailored to discrete or continuous data, further improving the fit.  The proposed framework naturally  converts the density comparison into a coefficient comparison, allowing standard inference tools (e.g., Wald tests) to be used. As such, the differential analysis amounts to performing density-level inference, provided the uncertainty in those estimates can be precisely accounted for.

The R codes and the data sets are available at  https://github.com/qz91/sef\_deDist/tree/main.

\section{Specially Designed Exponential Family for Density Estimation} 
Consider a set of observations $y_1, y_2, \ldots, y_m $ drawn independently from an unknown probability density function $f(y)$. In single cell applications, these are the library-size normalized gene expression counts in $m$ cells for a given individual. The seminal work of~\cite{efron96} introduced the specially designed exponential family~(SEF) as an effective approximation of $f(y)$, which takes the form 
$$
f_{\beta} (y) = \wh{\mu}_0(y) \exp(t(y)^\top \beta),
$$ 
where $\wh{\mu}_0(y)$ represents a nonparametric kernel density estimate~(KDE) based on the observed data and $t(y)$ denotes a vector of sufficient statistics, e.g., $t(y) = (1, y)$ or $t(y) = (1, y, y^2)$. The parameter value $\wh{\beta}$ is chosen by maximizing the likelihood, ensuring $t(y)$ moments of $f_{ \beta} (y)$ match their empirical averages  
$$
\int_{\mathcal{Y}} t(y) f_{ \beta} (y) = \frac{1}{n} \sum_{i=1}^n t(y_i). 
$$ 
Although directly fitting the SEF models may be computationally challenging, \cite{lindsey1974comparison,lindsey1974construction} suggested it could be rephrased in terms of Poisson regression model, facilitating efficient estimation via generalized linear models~(GLM). 

To make the problem computationally tractable, 
\cite{efron96} proposed to discretize the problem by the sample space $\mathcal{Y}$ into $K$ disjoint bins $\mathcal{Y}_k$, $\mathcal{Y} = \cup_{k=1}^K \mathcal{Y}_k$, and the data $(y_1, y_2, \ldots, y_m) \stackrel{\text{i.i.d.}}{\sim} f(y)$ are summarized to bin counts  
$$
s_k = \# \{y_i \in \mathcal{Y}_k \} \mbox{ for } k \in \{1, 2, \ldots, K\}. 
$$
We then model the bin counts via a Poisson regression model
\begin{align}\label{Poisson.model}
	s_k & \stackrel{\text{ind}}{\sim} \text{Poisson}\left( m \int_{\mathcal{Y}_k } \wh{\mu}_0(y ) \exp(t(y )^\top \beta ) dy \right ), \notag \\
	& \quad \approx \text{Poisson}\left( m |\mathcal{Y}_k| \wh{\mu}_0(y_{(k)}) \exp(t(y_{(k)})^\top \beta ) \right),  
\end{align}
where $y_{(k)}$ is a representative point for bin $\mathcal{Y}_k$ and $|\mathcal{Y}_k|$ is the bin size. Typically, $y_{(k)}$ is chosen as the midpoint of cell $\mathcal{Y}_k$.
With equal bin size $|\mathcal{Y}_1| = \ldots = |\mathcal{Y}_K|$,  the maximal likelihood estimation is equivalent  to solving
\begin{equation}\label{eqn:efron}
	X^\top [\vs - m \wh{\vmu}_0 \odot \exp(X\wh{\beta}) |\mathcal{Y}_K | ] = 0,  
\end{equation}
where $X \in \bbR^{K \times (p+1) }$ has $k$-th row being $(1, y_{(k)}, y_{(k)}^2, \ldots, y_{(k)}^p )$, $\vs  = (s_{1}, s_{2}, \ldots, s_{K}) \in \bbR^K$ represents the observed bin counts, and $\wh{\vmu}_0 = (\mu_0(y_{(1)}), \mu_0(y_{(2)}), \ldots, \mu_0(y_{(K)}) )$ is the estimated carrier density. As such,   $\wh{\vmu}_0 \odot \exp(X\wh{\beta}) $ provides a good density estimate up to a normalizing constant. This formulation allows efficient estimation using software such as  \texttt{glm()} in R.

\section{Modeling Multiple Samples from Single Cell RNA-sequencing Data}

In many modern biomedical applications, particularly in single-cell RNA-sequence (scRNA-seq) data, we observe gene expression levels for cells across multiple individuals. In this context, estimating densities for gene expression across different individuals and conditions becomes more interesting and challenging, especially when comparing distributions across groups, e.g., treatment vs. control. 

\subsection{Multi-sample Poisson Regression Model}

Before delving deeper, we introduce some notation here for better clarity. We consider a setup where the gene expression levels are modeled as $y_{icg}^{(t)} \sim f_{ig}^{(t)}(y)$, where $i$ indexes individuals, $c$ indexes cells, $g$ indexes genes, and $t$ indexes groups (e.g. $t_1$ for treatment, $t_2$ for control). For simplicity, we omit the index $g$, as we focus on individual-specific gene expression densities for a given gene.
We summarize the bin counts for each individual as: 
$$
\vs_i  = (s_{i1}, s_{i2}, \ldots, s_{iK}) \in \mathbb{N}^K, i \in \{1, 2, \ldots, n \}
$$ 
the vector of counts in $K$ bins $\mathcal{Y}_k$ as aforementioned, subject to $\sum_{k=1}^K s_{ik} = m_i$, where $m_i$ is the number of cells for individual $i$. 

The single-cell dataset can be modeled with observations from multiple individuals, and densities themselves exhibit variability. To capture this variability, we assume that each individual's density function, denoted as $f_i^{(t)}(y)$, is random with its associated probability measure $P_i^{(t)}$. Formally, we assume: 
\begin{ass}\label{ass:random}
	The density $f_i^{(t)}$ is associated with a random absolute continuous probability measure $P_{i}^{(t)} \in \mathcal{P}(\mathcal{Y})$, where $\mathcal{P}(\mathcal{Y})$ denote the space of probability measures defined on measurable space $(\mathcal{Y}, \mathcal{A})$. Here, $\mathcal{A}$ is the $\sigma$-algebra. We model 
	$
	P_{i}^{(t)} \stackrel{i.i.d}{\sim} \mathcal{D}_t,
	$
	where $\mathcal{D}_t \in \sP( \sP(\mathcal{Y}) ) $ is a  distribution over random probability measures. Additionally, we assume the existence of a mean probability measure $P^{(t)}$ with density $f^{(t)}$ such that 
	$$
	P^{(t)} (A) =  \Expect_{\mathcal{D}_t }[ P_i^{(t)} (A)], \quad  \forall A \in \mathcal{A}. 
	$$  
\end{ass} 

Assumption~\ref{ass:random} sets the foundation of our framework, where each individual's expression distribution is treated as an independent random object that fluctuates around the group-level mean density. The framework naturally covers the settings where densities themselves are random, particularly in hierarchical and empirical Bayes settings where the individual-level density $f_i^{(t)}$ arises from latent random effects or factors. Some examples include  Dirichlet processes, Gaussian random-effects and the Poisson-Gamma models that are commonly used for modeling single cell RNA-seq data \citep{sarkar2021separating}.

We extend the Poisson regression model \eqref{Poisson.model} and the corresponding score equation \eqref{eqn:efron}  to multiple samples
{
	\begin{equation}\label{eqn:single-density}
		X^\top[\vs_i - m_i M(\vs_1, \vs_2, \ldots, \vs_n ) \odot \exp(X\wh{\beta}^{t, i}) |\mathcal{Y}_K| ] = 0, \quad i=1,\cdots, n. 
	\end{equation} 
}
This time, we use a shared carrier density $\wh{\vmu}_0: = M(\vs_1, \vs_2, \ldots, \vs_n ) $, estimated from data from all the cells of all individuals via KDE. Moreover, we have an individual-specific modification factor $\exp(X\wh{\beta}^{t, i})$ that varies with $i$, allowing flexible adjustment per individual.

\begin{rem}[Adjustment of library sizes]
	{\rm 
		Unlike the original derivation of~\cite{efron96}, we explicitly integrate normalization factors such as   
		$m_i$ and $|\mathcal{Y}_K|$ in~\eqref{eqn:single-density}. In contrast, Efron's formulation omits these factors:
		\begin{equation}
			X^\top [\vs - \wh{\vmu}_0 \odot \exp(X\wh{\beta}) ] = 0.   
		\end{equation} 
		It does not fundamentally alter the estimation for the  parameters except the intercept term, which absorbs effects of factors like  $|\mathcal{Y}_K|$ and $m_i$. In other words, the intercept acts as a role of normalizing constant. We retain the explicit scaling for clarity and consistency in both theoretical analysis and empirical applications. 
	}
\end{rem}

\begin{rem}[Interpretation of coefficients] 
	{ \rm 
		The estimate $\wh{\beta}^{t,i}$ is the solution to the empirical score equation~\eqref{eqn:single-density}, which also maximize the Poisson likelihood. However, it is important to clarify its role. The coefficients $\beta^{t, i}$ are not intended to directly estimate the parameters in the density function $f_i^{(t)}(y) \propto f_0(y) \exp(t(y)^\top \gamma^{t,i})$.
		Instead, they serve as an adjustment  that maximizes the likelihood via matching the empirical moments. 
		To illustrate, consider the single-sample case. The empirical score function~\eqref{eqn:efron} rescaled by $m_i$ is given by:
		\begin{equation}
			X^\top [\vs_i/m_i -  \wh{\vmu}_0 \odot \exp(X\wh{\beta}^{t, i}) |\mathcal{Y}_K | ] = 0.
		\end{equation}
		Here, the carrier density $\wh{\vmu}_0 = M(\vs_i) $ is estimated from individual-level data. As $K$ and $m_i$ grow, 
		$\vs_{i,k}/m_i \rightarrow \int_{\mathcal{Y}_k} f_i^{(t)}(y) dy \approx f_i^{(t)}(y_{(k)} ) |\mathcal{Y}_k| $, and the KDE consistently estimates  $f_i^{(t)}(y)$. Thus $\beta^{t,i} \approx 0$ when  carrier density perfectly aligns with the individual density. This highlights that $\beta^{t,i}$ represents an adjustment factor rather than direct density parameters. 
	}
\end{rem}

\subsection{Group Comparison for Differential Density  Analysis} 
Suppose we compare two groups, treatment~($t_1$) and control~($t_2$), with sample sizes of $n_1$ and $n_2$ individuals  respectively, so that $n = n_1 + n_2$. 
The SEF fit for individual density corresponds to 
\begin{equation}\label{eqn:empirical-single}
	\wh{\vmu}_0 = M(\vs_1, \vs_2, \ldots, \vs_{n}),  \mbox{ and }  \wh{\beta}^{t, i}: X^\top \{ \vs_{i}/m_i - \wh{\vmu}_0 \odot \exp(X\wh{\beta}^{t,i}) |\mathcal{Y}_K |  \} = 0, \,  t = t_1, t_2,  
\end{equation} 
Meanwhile, regarding comparing two groups (e.g., treatment vs. control), we propose to use the following SEF framework for group-level density estimates:  
\begin{align}
	& X^\top \{ (\sum_{i=1}^{n_1} m_i)^{-1}\sum_{i=1}^{n_1} \vs_i - M(\vs_1, \ldots, \vs_{n} ) \odot \exp(X \wh{\beta}^{t_1} ) |\mathcal{Y}_K |\} = 0,    \label{eqn:empirical-M1} \\ 
	& X^\top \{ (\sum_{i=n_1+1}^{n} m_i)^{-1}\sum_{i=n_1+1}^{n_1+n_2} \vs_i - M (\vs_1, \ldots, \vs_{n} ) \odot \exp(X \wh{\beta}^{t_2} ) |\mathcal{Y}_K | \} = 0. \label{eqn:empirical-M2}
\end{align} 
Here, we have also estimated the carrier density by pooling all individuals' counts $(\vs_1, \vs_2, \ldots, \vs_n)$, representing the shared common carrier density of individuals across both groups. In single cell differential density analysis, this implies that we estimate the common carrier density using all the cells in the study. On top of that, the carrier density is modified by ``average effects'' $\wh{\beta}^{t_1}$ and $\wh{\beta}^{t_2}$ respectively, which characterize the group heterogeneity. More insights are provided by Proposition~\ref{prop:match}.

In the above discrete cases, we express $\wh{\vmu}_0$ as $M (\vs_1, \ldots, \vs_{n} )$, as it is a nonparametric density estimate given by some smoother $M(\cdot)$ applied to observed counts $\vs_1, \vs_2, \ldots, \vs_n$. 
Typically, the smoothing operation takes the form $M(\vs_i) = M \vs_i$, where $M$ is a $K\times K$ smoothing matrix. The pooled density estimate across all individuals is given by 
\begin{equation}\label{eqn:carrier}
	M(\vs_1, \vs_2, \ldots, \vs_n) = M(\sum_{i=1}^n \vs_i).
\end{equation}
This formulation ensures that $\wh{\vmu}_0$ captures the shared density structure across samples while allowing for individual- or group-specific adjustments via the exponential factors. 

As an important side note, the modification term $\exp(X \wh{\beta}^{t_1} )$ plays a role as  
$$ 
(\sum_{i \in T_1} m_i)^{-1} \sum_{i \in T_1} m_i \exp(X \wh{\beta}^{t_1, i}). $$  
In other words, the group-level modification term is a population-level summary of individual specific adjustments, weighted by the number of observations per individual. A similar relationship holds for group $t_2$. {
	Equations~\eqref{eqn:empirical-M1} and~\eqref{eqn:empirical-M2} are written in a rescaled form by dividing  the total cell counts 
	mainly for ease of theoretical analysis. In practice, implementation uses raw bin counts (e.g., $\sum_{i} \vs_i$ or $\vs_i$) without division, since Poisson regression requires integer-valued responses. As such, this gives identical results. 
} 

\subsection{Carrier Density: a Pooled Kernel Density Estimate}
In our model, we pool the gene expression counts over all the cells of $n$ individual samples and fit a kernel density estimator~(KDE), as in~\eqref{eqn: carrier}. For discrete Poisson regression model, this is precisely
\begin{equation}\label{eqn: carrier}
	\begin{aligned}
		& \wh{\vmu}_0 = \sum_{i=1}^{n} \frac{m_i}{\sum_{i=1}^{n} m_i} \sum_{j=1}^{m_i} \frac{ M_h(y | y_{(k)}) \b1\{y_{ij} \in \mathcal{Y}_k \} }{m_i} \\
		& = \sum_{i=1}^{n} w_i \sum_{k=1}^K \frac{M_h(y | y_{(k)}) s_{i,k} }{m_i}  =  M \cdot (\sum_{i=1}^n \vs_i)  
	\end{aligned}
\end{equation}
where $w_i = m_i/\sum_{i=1}^{n} m_i$, $M \in \bbR^K$ with $M_{ij} = (\sum_{i=1}^n m_i)^{-1} [M_h (y_{(i)} | y_{(j)})], i, j \in \{1, \ldots, K\}$, representing the kernel evaluated between discrete bins. From a continuous perspective when $K$ is sufficiently large, and $y_{ij}$ is chosen as a representative points for $\mathcal{Y}_k$, this estimator is expressed as: 
\begin{equation}\label{eqn:kde}
	\wh{\mu}_0(y) = \frac{1}{\sum_{i=1}^{n} m_i } \sum_{i=1}^{n} \sum_{j=1}^{m_i} K_h( y - y_{ij} ) = \sum_{i=1}^{n} \frac{m_i}{\sum_{i=1}^{n} m_i} \frac{1}{m_i} \sum_{j = 1 }^{m_i} K_h(y - y_{ij}) = \sum_{i=1}^{n} w_i \wh{f}_i (y), 
\end{equation}
where $\wh{f}_i (y)$ is the KDE for the $i$-th individual. This represents a weighted sum of KDEs across all individuals. 

In the remainder, we use $\Expect[ \cdot | f_i^{(t)}]$ to represent the expectation conditioned on the distribution $f_i^{(t)}$, i.e., when data follow the distribution with density $f_i^{(t)}$. If the density is determined by a latent variable $\vh_i^{(t)}$, this can  equivalently be written as $\Expect[ \cdot | \vh_i^{(t)} ]. $  

\begin{ass}\label{ass: moments}
	The following conditions hold: \begin{enumerate}
		\item For $d \in \{1, \ldots, p\}$, moment expectation 
		$ \Expect[ t_d (y) | f_i^{(t) }]  $
		exists, and  
		$$
		\Expect_{f^{(t)}} \{   t_d^2 (y)  \} < C_d < \infty, \forall t \in \{t_1, t_2\}. $$   
		\item Define the proportion of samples in each group $$
		r_{1, n} := \sum_{i\in T_1} w_i = \sum_{i\in T_1} m_i/(\sum_{i=1}^n m_i ) \mbox{ and } r_{2, n} := \sum_{i\in T_2} w_i = \sum_{i\in T_2} m_i/(\sum_{i=1}^n m_i ), $$
		we have  $r_{1, n} \rightarrow r_1, r_{2,n} \rightarrow r_2$ with $r_1 + r_2 = 1$. The weight sequence satisfies 
		$ \sum_{i=1}^n w_i^2 \rightarrow 0 $ as $n \rightarrow \infty$.   
		\item The kernel function used in fitting the KDE satisfies 
		$ \int_{\mathcal{Y}} M(s)ds = 1$, $\int_{\mathcal{Y}} s^2 M(s)ds = \mu_M < \infty $, and $\int_{\mathcal{Y}} M^2(s)ds < \infty$. 
	\end{enumerate}
\end{ass} 
If we further use $w_{1, i} = m_i/(\sum_{i \in T_1} m_i)$ and $w_{2, i} = m_i/(\sum_{i \in T_2} m_i)$, we have
$r_1^2 \sum_{i\in T_1} w_{1,i}^2 + r_2^2 \sum_{i\in T_2} w_{2,i}^2 = \sum_{i=1}^n w_i^2 \rightarrow 0$, which
implies $\sum_{i\in T_1} w_{1,i}^2 \rightarrow 0$, $\sum_{i\in T_2} w_{2,i}^2 \rightarrow 0$.

{
	Assumption~\ref{ass: moments} is relatively mild. It requires only finite-moment conditions for moments of sufficient statistics under consideration. It also assumes the cell counts in both groups maintain non-negligible portions of the total counts to ensure good estimation for densities in either group, which  prevents degenerated cases where a single sample's cell counts dominate, which makes density estimation infeasible. 
	The last assumption  is the standard condition in KDE theory. 
}

\begin{proposition}\label{prop:match}
	Under Assumptions~\ref{ass:random} and \ref{ass: moments} 
	our SEF framework using~\eqref{eqn:empirical-M1} and~\eqref{eqn:empirical-M2} asymptotically solves the system  
	\begin{equation}\label{eqn:asymp}
		\begin{aligned}
			& \int t(y) (r_1 f^{( t_1) } + r_2 f^{(t_2)})\exp(t(y)^\top \beta^{t_1}) dy = \int t(y)  f^{(t_1)}(y) dy, \\
			& \int t(y) (r_1 f^{(t_1)} + r_2 f^{(t_2)})\exp(t(y)^\top \beta^{t_2}) dy = \int t(y)  f^{(t_2)}(y) dy, 
		\end{aligned} 
	\end{equation} 
	where $\beta^{t_1}$ and $\beta^{t_2}$ match the moments of the pooled density with those of  $f^{t_1} $ and $f^{t_2}$, respectively. 
\end{proposition}
This proposition  leads to testing the discrepancy between two probability distributions with densities $f^{(t_1)}$ and $f^{(t_2)}$ via the hypothesis
$$
H_0: \int t(y)  f^{(t_1)}(y) dy = \int t(y)  f^{(t_2)}(y) dy, \quad  H_1: \int t(y)  f^{(t_1)}(y) dy \neq \int t(y)  f^{(t_2)}(y) dy.  
$$ 
More importantly, under the null hypothesis and given $r_1 + r_2 = 1$, we have $\beta^{t_1} = \beta^{t_2} = 0$. Otherwise,  $\beta^{t_1} \neq \beta^{t_2}$. Thus, testing density discrepancy is equivalent  to \textit{testing the difference of coefficient vectors}. Our proposed method simultaneously enables density estimation and group-wise discrepancy testing. 
In analysis of single cell data, this provides a test for differential expression distribution analysis between two groups $t_1$ and $t_2$. 
\begin{rem}
	{\rm 
		In the same vein, we find that the individual-level parameter $\beta^{t,i}$ satisfies 
		\begin{align*}
			\int t(y) (r_1 f^{( t_1) } + r_2 f^{(t_2)})\exp(t(y)^\top \beta^{t_1, i}) dy = \int t(y)  f_i^{(t_1)}(y) dy,  \\
			\int t(y) (r_1 f^{( t_1) } + r_2 f^{(t_2)})\exp(t(y)^\top \beta^{t_2,i }) dy = \int t(y)  f_i^{(t_1)}(y) dy, 
		\end{align*}
		In this sense, $\beta^{t,i}$ can be viewed as a random coefficient in parameter space $\Theta$, corresponding to the random density $f_i^{t}(y)$. More formally, we can define a distribution 
		$\beta^{t,i} \sim G_{\beta^{t}} \in \mathcal{P}(\Theta), $
		where $G_{\beta^{t}}$ is a distribution in the space of probability measures $\mathcal{P}(\Theta)$ defined on $\Theta$, such that 
		\begin{equation}\label{eqn:dist-beta}
			\begin{aligned}
				& \Expect_{G_{\beta^t }}[ \int_{\mathcal{Y}} t(y) (r_1 f^{( t_1) } + r_2 f^{(t_2)})  \exp(t(y)^\top \beta^{t_1, i})] dy \\
				& \quad =
				\int_{\mathcal{Y}} t(y) (r_1 f^{( t_1) } + r_2 f^{(t_2)}) \exp(t(y)^\top \beta^{t_1}) dy. 
			\end{aligned}     
		\end{equation}
		Their correspondence can be illustrate with the following mathematical diagram
		$$
		\begin{tikzcd}[row sep=large]
			f_i^{(t)} \sim \mathcal{D}_t \in \mathcal{P}(\mathcal{P}(\mathcal{Y}))  
			\arrow[r, shorten >= 5pt, shorten <= 5pt ] 
			\arrow[d, "\mathbb{E}_{\mathcal{D}_t}"'] 
			& \beta^{t,i} \sim G_{\beta^t} \in \mathcal{P}(\Theta) 
			\arrow[d, "\mathbb{E}_{G_{\beta^t}}"] \\
			f^{(t)}  
			\arrow[r, leftrightarrow, shorten >= 10pt, shorten <= 10pt ]  
			& \beta^{t} 
		\end{tikzcd}
		$$ 
		For a measure $\mathcal{P}_i^{(t)} \sim \mathcal{D}_t$ with density $f_i^{(t)}$, $t \in \{t_1, t_2\}$, there is an associated random parameter $\beta^{t, i}$ following distribution $G_{\beta^{t}}$ such that~\eqref{eqn:dist-beta} holds. 
	}
\end{rem}
\begin{rem}
	{\rm 
		We further remark that our framework naturally extends to multi-group comparison. Suppose we have three groups, each consisting of cells from different individuals. A carrier density $\wh{f}_0(y)$, representing the shared baseline structures, can be interpreted as an estimate for $\sum_{i=1}^3 r_i f^{(t_i)}(y)$, in light of Proposition~\ref{prop:match}. Based on this carrier, we can estimate the group-specific parameters $\wh{\beta}^{t_1}, \wh{\beta}^{t_2}, \wh{\beta}^{t_3}$, which capture the distinctions between density groups. 
		
		Importantly, our framework is model-agnostic: it does not require prior specification of distributional form. This flexibility arises from the combination of nonparametric KDE with a parametric adjustment, enabling accurate approximation of a broad range of probability densities while mitigating the risks of model misspecification. 
	}
\end{rem}

\section{Statistical Inference}
\subsection{Test Statistics based on SEF Estimates}
To test $H_0: \beta^{t_1} - {\beta}^{t_2}=0$, a natural test  statistic  is based on the SEF estimate 
$\wh{\beta}^{t_1} - \wh{\beta}^{t_2}$.  Theorem \ref{thm:cov} provides the approximate covariance of the SEF estimates. 
\begin{theorem}[Approximate covariance of SEF estimates]\label{thm:cov}
	Let $\wh{\beta}^{t, i}, \wh{\beta}^{t}, t = t_1, t_2 $, and their difference $\wh{\beta}^{t_1} - \wh{\beta}^{t_2} $ be the SEF estimators obtained from solving~\eqref{eqn:empirical-single},~\eqref{eqn:empirical-M1} and~\eqref{eqn:empirical-M2}. Their approximate covariance matrices are given as follows: 
	\begin{enumerate}
		\item For individual-level estimates $\wh{\beta}^{t, i}$, 
		$$
		\wh{\Sigma}_{\wh{\beta}^{t,i}} =  G_i^{-1} ( \sum_{j=1}^{n} Z_{ij}^\top \wh{\Cov}(\vs_j) Z_{ij} ) G_i^{-1}, 
		$$
		where $G_i = X^\top D( \wh{\vmu}_0 \odot \exp(X \wh{\beta}^{t,i})|\mathcal{Y}_k | ) X$, and  $Z^\top_{ij}  = X^\top [\delta_{ij} I_K - D(\exp(X\wh{\beta}^{t, i} )|\mathcal{Y}_k |) \frac{d\wh{\vmu}_0 }{d\vs_j  } ] $. 
		\item For group-level estimates $\wh{\beta}^{t_1}$,
		\begin{align*}
			& \wh{\Sigma}_{\wh{\beta}^{t_1}} =  G_{t_1}^{-1} (\sum_{j=1}^{n} Z_{t_1, j}^\top \wh{\Cov}(\vs_j) Z_{t_1, j}  ) G_{t_1}^{-1}, \quad   \wh{\Sigma}_{\wh{\beta}^{t_2}} =  G_{t_2}^{-1} (\sum_{j=1}^{n} Z_{t_2, j}^\top \wh{\Cov}(\vs_j) Z_{t_2, j}  ) G_{t_2}^{-1}.
		\end{align*}
		
		\item For group-level difference $\wh{\beta}^{t_1} - \wh{\beta}^{t_2} $, 
		$$
		\wh{\Sigma}_{\wh{\beta}^{t_1} - \wh{\beta}^{t_2} } = \sum_{j=1}^n ( G_{t_1}^{-1} Z_{t_1, j}^\top - G_{t_2}^{-1} Z_{t_2, j}^\top ) \wh{\Cov}(s_j) (G_{t_1}^{-1} Z_{t_1, j}^\top - G_{t_2}^{-1} Z_{t_2, j}^\top )^\top.   
		$$ 
	\end{enumerate}
	In the above, we denote
	\begin{equation}\label{eqn:mat-G} 
		G_{t_1} = X^\top D( \wh{\vmu}_0 \odot \exp(X\wh{\beta}^{t_1})|\mathcal{Y}_k |)  X,  \quad  G_{t_2} = X^\top D(  \wh{\vmu}_0 \odot \exp(X\wh{\beta}^{t_2}) |\mathcal{Y}_k | )X, 
	\end{equation}
	and 
	\begin{align*}
		& Z_{t_1, j}^\top = X^\top \left\{ \frac{\mathbf{1}\{j \in T_1\}}{ \sum_{i \in T_1} m_i } I_K  - D(\exp(X\wh{\beta}^{t_1}))|\mathcal{Y}_k |) d\wh{\vmu}_0/d\vs_j \right\}, \\
		& Z_{t_2, j}^\top = X^\top  \left\{ \frac{\mathbf{1}\{j \in T_2\}}{\sum_{i \in T_2} m_i}   I_K - D(\exp(X\wh{\beta}^{t_2}) |\mathcal{Y}_k | )) d\wh{\vmu}_0/d\vs_j \right\}.
	\end{align*}  
\end{theorem}
\begin{corollary}
	If further the carrier density satisfies $d\wh{\vmu}_0/d\vs_j  = M$ for some matrix $M \in \bbR^{K \times K}$, then the approximate covariance matrices simplify as follows, 
	\begin{align*}
		&  \wh{\Sigma}_{\wh{\beta}^{t_1}} =  G_{t_1}^{-1} ( Z_{t_1, t_1}^{\top} \sum_{j \in T_1}  \wh{\Cov}(\vs_j) Z_{t_1, t_1} + Z_{t_1, t_2}^{\top} \sum_{j \in T_2}  \wh{\Cov}(\vs_j) Z_{t_1, t_2}  ) G_{t_1}^{-1}, \\ 
		& \wh{\Sigma}_{\wh{\beta}^{t_1} - \wh{\beta}^{t_2}} = ( G_{t_1}^{-1} Z_{t_1, t_1}^\top - G_{t_2}^{-1} Z_{t_2, t_1}^\top ) \wh{\Cov}( \sum_{j\in T_1} s_j) ( G_{t_1}^{-1} Z_{t_1, t_1}^\top - G_{t_2}^{-1} Z_{t_2, t_1}^\top )^\top \\
		& \quad + ( G_{t_1}^{-1} Z_{t_1, t_2}^\top - G_{t_2}^{-1} Z_{t_2, t_2}^\top ) \wh{\Cov}( \sum_{j\in T_2} s_j) ( G_{t_1}^{-1} Z_{t_1, t_2}^\top - G_{t_2}^{-1} Z_{t_2, t_2}^\top )^\top, 
	\end{align*}  
	where the simplified weight matrices are: 
	\begin{align*}
		&  Z_{t_1, t_1}^{\top} = X^\top \left[ \frac{I_K}{\sum_{i \in T_1} m_i}  - D(\exp(X\wh{\beta}^{t_1})|\mathcal{Y}_k |) M \right], Z_{t_1, t_2}^{\top} = X^\top \left[ -D(\exp(X\wh{\beta}^{t_1}) |\mathcal{Y}_k |) M \right], \\
		&  Z_{t_2, t_1}^{\top} = X^\top \left[  - D(\exp(X\wh{\beta}^{t_2})|\mathcal{Y}_k |) M \right], Z_{t_2, t_2}^{\top} = X^\top \left[ \frac{I_K}{\sum_{i \in T_2} m_i}  -D(\exp(X\wh{\beta}^{t_2})|\mathcal{Y}_k |) M \right]. 
	\end{align*}
	A good estimator for $\Cov(\sum_{i \in T_1} \vs_i)$ is 
	\begin{align*}
		\wh{\Cov}(\sum_{i \in T_1} \vs_i ) & =  \sum_{i \in T_1} \left( D(\vs_i) - 2 \frac{\vs_i\vs_i^\top}{m_i} + \frac{\vs_i}{m_i} \frac{\vs_i^\top}{m_i} \right) \\
		& + \sum_{i \in T_1} m_i^2 \left( \frac{\vs_i}{m_i} -  \frac{\sum_{i \in T_1} \vs_i }{ \sum_{i \in T_1} m_i  } \right) \left( \frac{\vs_i}{m_i} - \frac{\sum_{i \in T_1} \vs_i}{ \sum_{i \in T_1}  m_i} \right)^\top,
	\end{align*}
	and similarly for  $\Cov(\sum_{i \in T_2} \vs_i)$. 
\end{corollary}

Given the estimated covariance matrix $\wh{\Sigma}$ (subscripts suppressed for brevity),
inference on the group-level difference $\wh{\beta}^{t_1} - \wh{\beta}^{t_2}$ follows from 
$
\wh{\Sigma}^{-1/2} (\wh{\beta}^{t_1} - \wh{\beta}^{t_2}) \stackrel{d}{\longrightarrow} \mathcal{N}(0, I_{p+1}),
$
which leads to the joint Wald test 
$$
(\wh{\beta}^{t_1} - \wh{\beta}^{t_2})^\top \wh{\Sigma}^{-1} (\wh{\beta}^{t_1} - \wh{\beta}^{t_2}) > \chi_{p+1,1-\alpha}^2,
$$
where $\chi_{p+1,1-\alpha}^2$ represents the $(1-\alpha)$ quantile of the $\chi^2$ distribution with $p+1$ degrees of freedom.

{As an alternative, 
	we can separate $\beta^{t} = (\beta_0^{t}, \beta_1^{t})$ with $\beta_1^t \in \bbR^p, t = t_1 , t_2$ and  conduct inference on the non-intercept component $(\wh{\beta}^{t_1}_1 - \wh{\beta}^{t_2}_1)$. 
	In fact, $\beta_1^t \in \bbR^p$ alone suffices to determine the underlying distribution (see Section~\ref{apxsec:equivalent} of Supplementary Material for detailed dicussion), and is the focus of earlier work such as~\cite{efron96}.  In particular, we consider the marginal Wald test for the subvector 
	\begin{equation}\label{eqn:test-beta1}
		(\wh{\beta}^{t_1}_1 - \wh{\beta}^{t_2}_1 )^\top \wh{\Sigma}_{1,1}^{-1}(\wh{\beta}^{t_1}_1 - \wh{\beta}^{t_2}_1)  > \chi_{p, 1-\alpha}^2,  
	\end{equation}
	where $\wh{\Sigma}_{1,1}\in \bbR^{p\times p}$ is the submatrix of $\wh{\Sigma}$ corresponding to the non-intercept components. This test, called the SEF test in the remaining of the paper, is adopted in both simulations and real-data analysis.

	The subvector-based test is often more favorable. First, the normalization constant $\sum_{i=1}^{n_1} m_i$, $|\mathcal{Y}_K|$ may be misspecified, which  does not affect the parameters $\wh{\beta}_1^{t}$ but can shift the intercept $\wh{\beta}_0^t$. Second, one may deliberately avoid explicit normalization, as in~\cite{efron96}, treating intercept as a free constant to ensure a valid density. Thus, $\beta_1^t$ captures the distributional differences and is the appropriate target for inference.  Focusing on $\beta_1^{t}$ therefore avoids artifacts due to normalization while aligning with prior methodological practice.
	
	
	We emphasize that the proposed test~\eqref{eqn:test-beta1} is exactly equivalent to performing test on $\wh{\beta}_1^{t_1} - \wh{\beta}_1^{t_2} $ obtained from the centered score equations below (see Section~\ref{apxsec:equivalent} of Supplementary Material details),  
	\begin{equation}\label{eqn:no-intecept}
		\begin{aligned}
			& X_{t_1}^\top \left\{ \frac{\sum_{i\in T_1} \vs_i }{ \sum_{i\in T_1} m_i} -   \wh{\vmu}_0 \odot \exp(  \wh{\beta}^{t_1}_0 + X_{t_1}\wh{\beta}^{t_1}_1  )  |\mathcal{Y}_K | \right\} = 0, \\
			& X_{t_2}^\top \left\{ \frac{\sum_{i\in T_2} \vs_i }{ \sum_{i\in T_2} m_i} -   \wh{\vmu}_0 \odot \exp(  \wh{\beta}^{t_2}_0 + X_{t_2}\wh{\beta}^{t_2}_1  )|\mathcal{Y}_K |  \right\} = 0, \\ 
			& \mbox{ sub to }  \sum_{k=1}^K   \wh{\mu}_0(y_{(k)}) \exp(  \wh{\beta}^{t}_0 + (t_c(y_{(k)}) - \bar{T}_c^{t})^\top \wh{\beta}^{t}_1  )|\mathcal{Y}_K |  = 1, t \in \{t_1, t_2\}. \\
		\end{aligned}
	\end{equation}
	Here $X_{t}^\top \in \bbR^{p \times K}, t \in \{t_1, t_2\}$ has $k$-th column corresponding to the centered transformation of sufficient statistics: $t_c(y_{(k)}) - \bar{T}_c^{t}$, where $t_c(y) = \{y, y^2, \ldots, y^{p}\}$ excludes the intercept. The sample average
	\begin{equation}
		\bar{T}_c^{t_1} = \frac{1}{\sum_{i \in T_1} m_i} \sum_{i \in T_1} \sum_{j=1}^{m_i} t_c(y_{ij}). 
	\end{equation}
	is computed from raw expression counts across all cells in group $T_1$, and the same can be done for group $T_2$. This centered formulation allows  direct computation for covariance matrix for $\wh{\beta}_1^{t_1} - \wh{\beta}_1^{t_2}$ according to formula~\eqref{eqn:asymp-emp-cov-beta1} in Corollary~\ref{cor:no-intercept}, by passing the need to compute and then reduce the full $(p+1)\times (p+1)$ covariance matrix.

\begin{rem}
	{\rm 
		We remark on several extensions. While the original approach primarily matches the raw~(non-centered) moments, it is straightforward to adapt the method for central moments by replacing the sufficient statistics with 
		$\wt{t}(y) = (1, y-\mu, (y-\mu)^2, \ldots, (y-\mu)^p )$, 
		where $\mu$ can be estimated at individual level by $\sum_{j=1} y_{ij}/m_i$ or $\sum_{k=1}^K s_{i,k} y_{(k)}/m_i$ for~\eqref{eqn:empirical-single}. In group-wise setting, an appropriate estimate is 
		$\sum_{k=1}^K (\sum_{i \in T_1} s_{i, k}) y_{(k)}/(\sum_{i\in T_1} m_i)$ for~\eqref{eqn:empirical-M1}. The resulting design matrix $\tilde{X}$ contains rows of these centered sufficient statistics. 
	} 
\end{rem} 
By adjusting the choice of moments, the SEF framework can target different types of distributional discrepancies. For instance, if two densities share the same first and second moments, SEF with $t(y) = (1, y, y^2)$ is incapable of distinguishing them; we can then expand the sufficient statistics to $t(y) = (1, y, y^2, y^3)$ to detect higher-order differences. If the scientific interest is in a single moment, such as the $p$-th moment, one may simply use $t(y) = (1, y^p)$. 
\medskip

In real data application,  we recommend starting with a small $p$ (e.g., $p=2-3$) to avoid over-fitting of the data, especially when the number of cells is not too large.  After estimation, compute the $t$-values for each coefficient in $\wh{\Sigma}^{-1/2} \wh{\beta}^{t_1}$ and $\wh{\Sigma}^{-1/2} \wh{\beta}^{t_2}$ separately. The intercept $\wh{\beta}_0^{t}$ serves only for normalization, and its value is of no statistical interest. We retain only those moment terms that are significant for at least one group; otherwise, the total relative variance (TRV) criterion of~\cite{efron96} (Section 5) can be used for model selection. 

\subsection{Asymptotic Properties}
In this section, we establish the asymptotic properties of the estimates derived from previous sections. Specifically, we analyze their consistency and asymptotic normality. 
Notably, solving~\eqref{eqn:empirical-M1} 
when $K$ is sufficiently large corresponds to maximize 
$$
B_{n}^{t_1}(\beta) := -\log \wh{c}(\beta) + \beta^\top \bar{T}^{t_1} = -\log \wh{c}(\beta) + \beta^\top \frac{\sum_{i \in T_1} \sum_{j=1}^{m_i} t(y_{ij})}{\sum_{i \in T_1} m_i}, 
$$
where 
\begin{equation}\label{eqn:chat}
	\wh{c}(\beta) = \int_{\mathcal{Y}} \wh{\mu}_0(y)  \exp(t(y)^\top \beta) dy.
\end{equation} 
For further analysis, we define
\begin{align*}
	& A_{n}^{t_1}(\beta) = -\log \wh{c}(\beta) + \beta^\top \xi^{t_1},  \quad  A^{t_1}(\beta) = -\log c(\beta) + \beta^\top \xi^{t_1},   
\end{align*} 
where 
\begin{equation}\label{eqn:c}
	c(\beta) = \int_{\mathcal{Y}}  (r_1 f^{t_1} + r_2 f^{t_2}) \exp(t(y)^\top \beta),
\end{equation}
and $\xi^{t_1} = \int t(y) f^{(t_1)} (y) dy$.
Similarly, we define $ B_{n}^{t_2}(\beta),  A_{n}^{t_2}(\beta)$, and $A^{t_2}(\beta)$ accordingly.  

To establish the consistency, we impose the following assumption:  
\begin{ass}\label{ass:consistency}
	For  $\forall t \in \{t_1, t_2\}$, we assume 
	(i) {Uniform convergence:} $\sup_{\beta \in \Theta} | B^{t}_n(\beta) - A^{t} (\beta) | \stackrel{p}{\longrightarrow} 0$. (ii) {Strong concavity:}  $A^{t}(\beta^{t}) > \sup_{\|\beta - \beta^{t}\| \ge \epsilon } A^{t}(\beta)$. 
\end{ass}
Assumption~\ref{ass:consistency} is standard in proving the consistency, as
$ \sup_{\beta \in \Theta} | B^{t}_n(\beta) - A_n^{t} (\beta) | \rightarrow^p 0$ is a well-known condition in the literature to prove consistency while $ A_n^{t} (\beta) \stackrel{p}{\rightarrow}  A^{t} (\beta) $ follows from the convergence of KDE, as mentioned earlier. 
Specifically, we have $\wh{c}(\beta) - c(\beta) = o_p(1)$ and $\log( \wh{c}(\beta) ) - \log( c(\beta) )= o_p(1)$ by the continuous mapping theorem. Consequently, we obtain
$
\wh{\beta}^{t} \stackrel{p}{\longrightarrow} \beta^{t}, t \in \{ t_1, t_2\}. 
$ 
\begin{theorem}\label{thm:consistency}
	Under Assumptions~\ref{ass:random}-\ref{ass:consistency}, the parameter estimates $\wh{\beta}^{t}, t \in \{t_1, t_2\}$ are consistent, i.e., 
	$
	\wh{\beta}^{t} \stackrel{p}{\longrightarrow} \beta^{t}  
	$
	for  $t \in \{t_1, t_2\},$ 
	as $\sum_{i=1}^n w_i^2 \rightarrow 0$. 
\end{theorem} 
Next, we establish the asymptotic  covariance structure and normality of the estimators.

\begin{theorem}[Asymptotic Normality]\label{thm:asymp}
	Under Assumptions~\ref{ass:random}-\ref{ass:consistency}, the asymptotic covariance matrix for $( \wh{\beta}^{t_1} - \wh{\beta}^{t_2}) - (\beta^{t_1} - \beta^{t_2} ) $ is 
	\begin{equation}\label{eqn:asymp-cov}
		\begin{aligned}
			& \Sigma_{\wh{\beta}^{t_1} - \wh{\beta}^{t_2} }   = H_{\beta^{t_1}}^{-1} \Cov(\bar{T}^{t_1} - \frac{\partial \log \wh{c}(\beta^{t_1})}{\partial \beta} )  H_{\beta^{t_1}}^{-1}  + H_{\beta^{t_2}}^{-1} \Cov(\bar{T}^{t_2} - \frac{\partial \log \wh{c}(\beta^{t_2})}{\partial \beta} )  H_{\beta^{t_2}}^{-1} \\ 
			& \quad + H_{\beta^{t_1}}^{-1} \Cov(\bar{T}^{t_1} - \frac{\partial \log \wh{c}(\beta^{t_1})}{\partial \beta},\bar{T}^{t_2} - \frac{\partial \log \wh{c}(\beta^{t_2})}{\partial \beta} ) H_{\beta^{t_2}}^{-1} \\
			& \quad + H_{\beta^{t_2}}^{-1} \Cov(\bar{T}^{t_2} - \frac{\partial \log \wh{c}(\beta^{t_2})}{\partial \beta}, \bar{T}^{t_1} - \frac{\partial \log \wh{c}(\beta^{t_1})}{\partial \beta}) H_{\beta^{t_1}}^{-1}.    
		\end{aligned}
	\end{equation} 
	Thus, 
	$$
	\Sigma_{\wh{\beta}^{t_1} - \wh{\beta}^{t_2} }^{-1/2} \{( \wh{\beta}^{t_1} - \wh{\beta}^{t_2}) - (\beta^{t_1 } - \beta^{t_2 } ) \} \stackrel{d}{\rightarrow} \mathcal{N}(0, I_{p+1}). 
	$$
\end{theorem} 
The following theorem ensures that the empirical covariance matrix converges to its population counterpart, thereby validating its use in inferential procedures. 
\begin{theorem}[Convergence of empirical covariance]\label{thm:cov-converge}
	Under the same conditions as previous theorems, the empirical covariance matrix $\wh{\Sigma}_{\wh{\beta}^{t_1} - \wh{\beta}^{t_2} }$ converges in probability to $\Sigma_{\wh{\beta}^{t_1} - \wh{\beta}^{t_2} } $ given in~\eqref{eqn:asymp-cov}. That is, 
	$ \wh{\Sigma}_{\wh{\beta}^{t_1} - \wh{\beta}^{t_2} } \stackrel{p}{\rightarrow} \Sigma_{\wh{\beta}^{t_1} - \wh{\beta}^{t_2} }. $ 
	Consequently, we obtain the asymptotic normality result:
	$$
	\wh{\Sigma}_{\wh{\beta}^{t_1} - \wh{\beta}^{t_2} }^{-1/2} \{  (\wh{\beta}^{t_1} - \wh{\beta}^{t_2}) - ( \beta^{t_1} - \beta^{t_2})  \}  \stackrel{d}{\rightarrow } \mathcal{N}(0, I_{p+1}). 
	$$
\end{theorem}

\begin{corollary}[Inference without intercepts]\label{cor:no-intercept}
	Suppose $\wh{\beta}^{t_1}_1, \wh{\beta}^{t_2}_1$ are solutions to the equation~\eqref{eqn:no-intecept}. Their difference  $\wh{\beta}^{t_1}_1 - \wh{\beta}^{t_2}_1$ has an estimated covariance 
	\begin{equation}\label{eqn:asymp-emp-cov-beta1}
		\begin{aligned}
			& \wh{\Sigma}_{\wh{\beta}^{t_1}_1 - \wh{\beta}^{t_2}_1 } = ( \wt{G}_{t_1}^{-1} \wt{Z}_{t_1, t_1}^\top - \wt{G}_{t_2}^{-1} \wt{Z}_{t_2, t_1}^\top ) \wh{\Cov}( \sum_{j\in T_1} s_j) ( \wt{G}_{t_1}^{-1} \wt{Z}_{t_1, t_1}^\top - \wt{G}_{t_2}^{-1} \wt{Z}_{t_2, t_1}^\top )^\top \\
			& \quad + ( \wt{G}_{t_1}^{-1} \wt{Z}_{t_1, t_2}^\top - \wt{G}_{t_2}^{-1} \wt{Z}_{t_2, t_2}^\top ) \wh{\Cov}( \sum_{j\in T_2} s_j) ( \wt{G}_{t_1}^{-1} \wt{Z}_{t_1, t_2}^\top - \wt{G}_{t_2}^{-1} \wt{Z}_{t_2, t_2}^\top )^\top,  
		\end{aligned}
	\end{equation}
	where 
	\begin{align*}
		& \wt{Z}_{t_1, t_1}^{\top} = X_{t_1}^\top \left[ \frac{1}{\sum_{i \in T_1} m_i} I_K - D(\exp(\wh{\beta }^{t_1}_0  + X_{t_1}\wh{\beta}^{t_1}_1) |\mathcal{Y}_K| )  M \right], \\
		& \wt{G}_{t_1} = X_{t_1}^\top D( \wh{\vmu}_0 \odot \exp(\exp(\wh{\beta}^{t_1}_0  + X_{t_1}\wh{\beta}^{t_1}_1))|\mathcal{Y}_K|)  X_{t_1},  \wt{Z}_{t_1, t_2}^{\top} = X_{t_1}^\top \left[ -D(\exp(\wh{\beta}^{t_1}_0  + X_{t_1}\wh{\beta}^{t_1}_1) |\mathcal{Y}_K| ) M \right]  
	\end{align*}
	The corresponding asymptotic structure is given by
	\begin{equation}  \label{eqn:asymp-cov-beta1}
		\begin{aligned}
			& \Sigma_{ \wh{\beta}_1^{t_1} - \beta_1^{t_2}} = \wt{H}_{\beta^{t_1}}^{-1} \Cov(\bar{T}_c^{t_1} - \frac{\partial \log \wh{c}(\beta^{t_1})  }{\partial \beta_1 }  )  \wt{H}_{\beta^{t_1}}^{-1}  + \wt{H}_{\beta^{t_2}}^{-1} \Cov(\bar{T}_c^{t_2} - \frac{\partial \log \wh{c}(\beta^{t_2})  }{\partial \beta_1 } )  \wt{H}_{\beta^{t_2}}^{-1} \\ 
			& \quad + \wt{H}_{\beta^{t_1}}^{-1} \Cov\left(\bar{T}_c^{t_1} - \frac{\partial \log \wh{c}(\beta^{t_1})  }{\partial \beta_1 } ),\bar{T}_c^{t_2} - \frac{\partial \log \wh{c}(\beta^{t_2})  }{\partial \beta_1 } \right) \wt{H}_{\beta^{t_2}}^{-1} \\
			& \quad + \wt{H}_{\beta^{t_2}}^{-1} \Cov\left(\bar{T}_c^{t_2} - \frac{\partial \log \wh{c}(\beta^{t_2})  }{\partial \beta_1 }, \bar{T}_c^{t_1} - \frac{\partial \log \wh{c}(\beta^{t_1})  }{\partial \beta_1 } \right) \wt{H}_{\beta^{t_1}}^{-1}, \\
		\end{aligned}
	\end{equation}  
	where $ \wt{H}_{\beta} = \frac{\partial \log c(\beta)}{\partial \beta_1 \partial \beta_1^{\top}} $. 
	Consequently, 
	$$
	\wh{\Sigma}_{\wh{\beta}^{t_1}_1 - \wh{\beta}^{t_2}_1 }^{-1/2} \{ (\wh{\beta}^{t_1}_1 - \wh{\beta}^{t_2}_1) -  (\beta^{t_1}_1 - \beta^{t_2}_1) \} \stackrel{d}{\longrightarrow} \mathcal{N}(0, I_p). 
	$$
\end{corollary} 
This final result ensures that inference on the non-intercept terms follows a well-defined asymptotic distribution, allowing for statistical testing of shape differences between two distributions. 

\begin{rem}
	{\rm 
		Our proposed framework is universal in the sense that it requires no prior specification of the underlying model and achieves good approximation by combining nonparametric with parametric exponential tilting. This flexibility comes at the cost of a slower convergence rate of $\wh{\beta}^{t_1} - \wh{\beta}^{t_2}$ compared to the usual $n^{-1/2}$ rate attained by parametric MLE. Because the KDE step introduces both bias and additional variance, one must balance the bandwidth choice to control the trade-off. In our setting, we can set the bandwidth as $h = O( (\sum_{i=1}^n w_i^2)^{\frac{1}{5} + 2\delta} )$ (small $\delta > 0$), ensuring that the scaled bias is negligible in the CLT. This yield the convergence rate $(\sum_{i=1}^n w_i^2)^{-\frac{2}{5} + \delta}$, where $1/(\sum_{i=1}^n w_i^2)$ acts as effective sample size. When $m_i \equiv m$, the rate attains its fastest order $n^{-\frac{2}{5}+\delta}$. Full technical details are provided in Section~\ref{apxsec:asymp} of the Supplementary. 
	} 
\end{rem}

\section{Simulation Study}
We conduct  simulation studies to evaluate the performance of our proposed method under various data distribution settings. Within each one, we consider several test cases to assess the null distribution and the power  of the SEF  test in detecting distributional differences between two groups. To assess statistical power, we compute the empirical rejection rate at a nominal Type I error level of $\alpha  = 0.05$.
We compare the proposed method~(SEF) to the following competing methods: (i) The  generic method of moments~(MoM). We construct the test statistic: 
$
( \bar{T}^{t_1} - \bar{T}^{t_2} )^{\top} [\Cov(\bar{T}^{t_1}) +  \Cov(\bar{T}^{t_2}) ]^{-1} ( \bar{T}^{t_1} - \bar{T}^{t_2} ) >  \chi_{p, 1 - \alpha}^2; 
$  
(ii) The pseudo-bulk two-sample t-test. (iii) The pseudo-bulk non-parametric Komolgorov-Smirnov goodness-of-fit test.

We include pseudo-bulk, aggregation-based methods in lieu of our single cell RNA-sequencing real data application.
Throughout all simulation experiments and downstream analyses, we employ the SEF estimator excluding the effect of the intercept normalizing constant~(see Corollary~\ref{cor:no-intercept}). Previous experiments including the intercept normalizing constant yielded identical results, so we only report the version as in ~\eqref{cor:no-intercept} to avoid redundancy. 

\subsection{Simulation Settings}  

We assess the robustness of the SEF test under a range of distributional settings. In each setting, we simulate data for two groups $(t = 1, 2)$ with $n_1 = n_2 = 100$ individuals per group. For individual $i$ in group $t$, the cell number is generated from the discrete uniform distribution
$
m_i \sim \textrm{Unif}\{300, 301, \ldots, 1000\}, 
$
mimicking more realistic scenarios of varying cell counts across individuals. 
We investigate the following models that have been shown to be appropriate for modeling scRNA-seq count data \citep{sarkar2021separating}:
\begin{enumerate}
	\item {Poisson-Gamma mixture model. } We generate 
	$$
	y_{ij} \sim \text{Poisson}(\lambda_i^{(t)}), \lambda_i^{(t)} \sim \mbox{Gamma}(\gamma_t, \eta_t),
	$$
	where $\gamma_t, \eta_t$ are the shape and rate parameters, respectively. The marginal distribution is 
	$$
	f^{(t)}(y) = \frac{\Gamma(y + \gamma_t )}{\Gamma(\gamma_t) y !} (\frac{\eta_t}{1+\eta_t})^{\gamma_t} (\frac{1}{1+\eta_t})^{y},
	$$
	which is a negative binomial $\text{NB}(\gamma, \eta/(1+\eta))$ where $\gamma_t$ is an integer. 
	We consider both mean and variance shift settings. For mean shift, we  fix the variance $\sigma_t^2 = \gamma_t/\eta_t + \gamma_t/\eta_t^2 = 15$ for both groups. Set $\mu_1 = \gamma_1/\eta_1 = 10$ and vary $\mu_2$ from $10$ to $12$ with $15$ equally spaced values. 
	For  variance shift setting,  we fix $\mu_1 = \mu_2 = 10$ and $\sigma^{(1)2} = 15$, while varying $\sigma^{(2)2}$ from $15$ to $23$ in $10$ equally spaced values. 
	The carrier density is estimated using a discrete Gaussian kernel.
	\item {Zero-inflated negative binomial (ZINB) model.} To mimic sparser cases of single-cell RNA-seq data, we generate  a zero-inflated negative binomial model.
	$$
	y_{ij}^{(t)} \sim \mbox{ZINB}(\mu_{i}^{(t)}, \theta^{(t)}, \pi^{(t)}), \mu_{i}^{(t)} \sim \textrm{truncNorm}(\mu^{(t)},\sigma^{(t),2},0)
	$$
	where $\mu^{(t)}$ is the mean, $\theta^{(t)}$ is the dispersion parameter, and $\pi^{(t)}$ is the zero-inflation probability for group $t$. 
	We first condition the setting where $\mu^{(t_1)}$ and $\mu^{(t_2)}$ are different, which corresponds to both mean and variance shift between the two groups. 
	We also consider the setting  of  differences in variance by varying the dispersion parameter: 
	$
	\theta^{(2)} = b_{\theta} \theta^{(1)},
	$
	with $\mu^{(1)} = \mu^{(2)} = 10$, $\sigma^{(1),2} = \sigma^{(2),2} =0.1$ and $\pi^{(1)} = \pi^{(2)} = 0.5$. In the baseline setting, $\theta^{(1)} = \theta^{(2)} = 10$, and $m_i$ follows the same distribution in prior models. The carrier density is again estimated with a discrete Gaussian kernel.
\end{enumerate}

\subsection{Simulation Results} 
The reliability of our SEF test is demonstrated by the QQ-plots the negative log-transformed p-values generated for the Poisson-Gamma, and zero-inflated models, in the scenario where both groups have identical distributions and therefore equivalent parameters. Our observed p-values approximately follow the uniform distribution as expected under equivalent distribution scenarios for all models (Figure \ref{fig:simQQplot} top panel). 

\begin{figure}
	\centering
	\includegraphics[width = 0.45\textwidth, height=0.3\textheight]{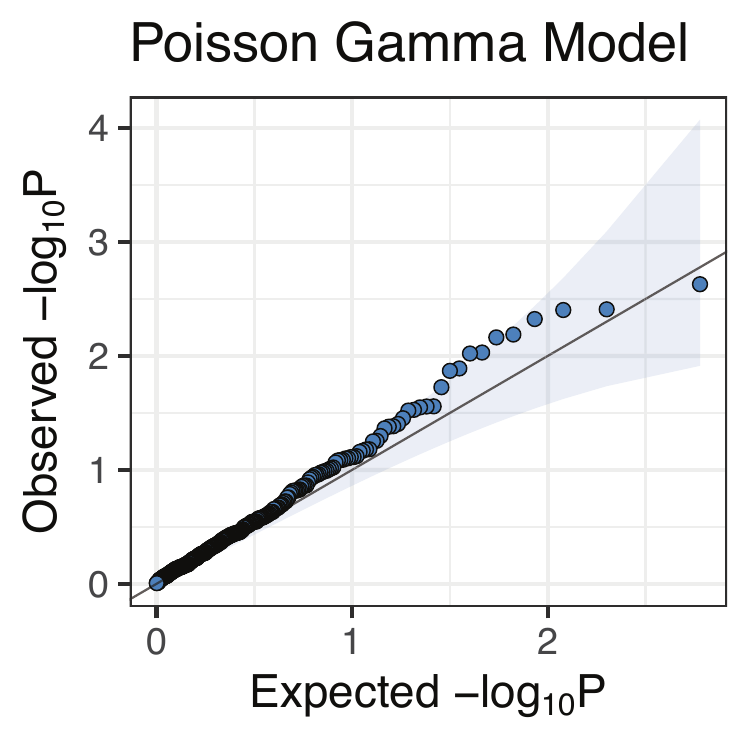} 
	\includegraphics[width = 0.45\textwidth,height=0.3\textheight]{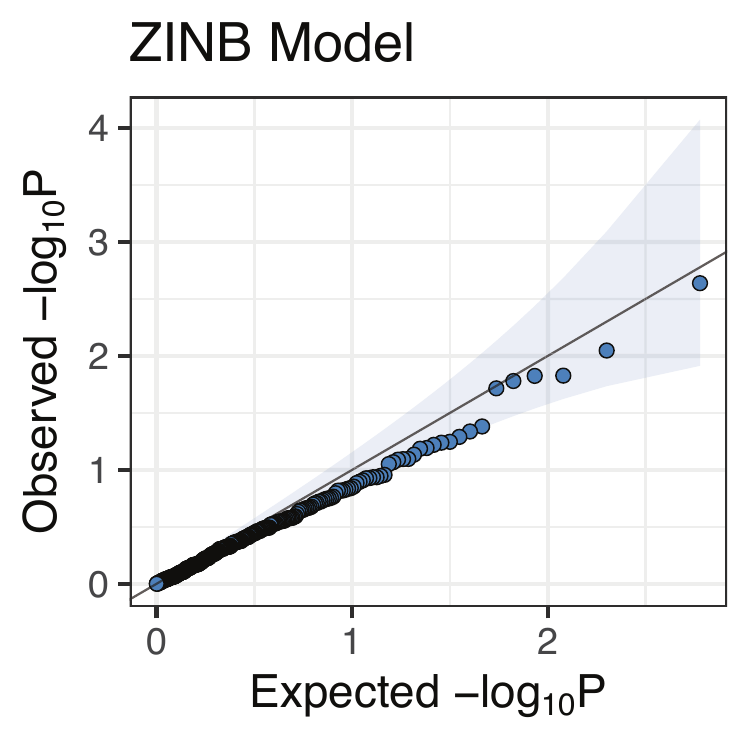} 
	\includegraphics[width = 0.45\textwidth,height=0.3\textheight]{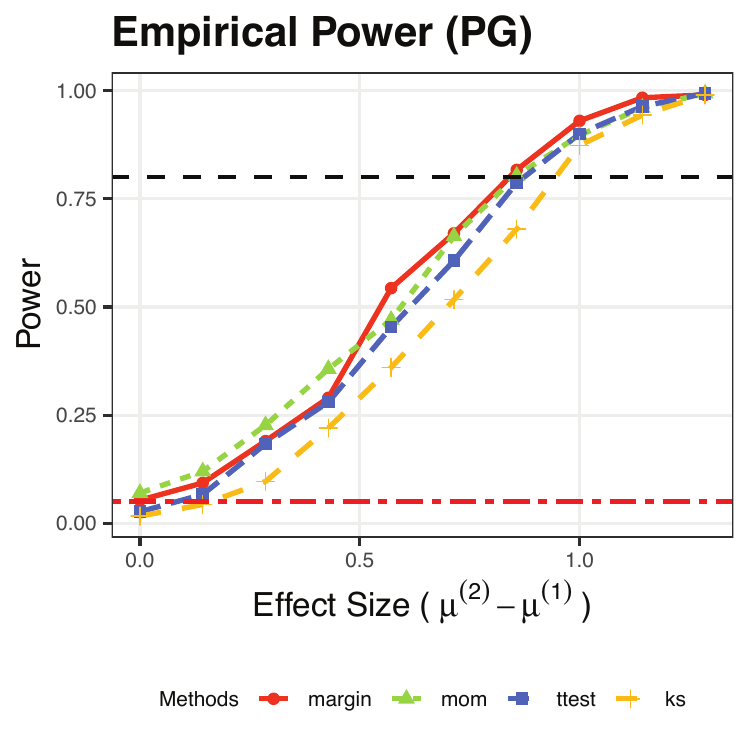} 
	\includegraphics[width = 0.45\textwidth,height=0.3\textheight]{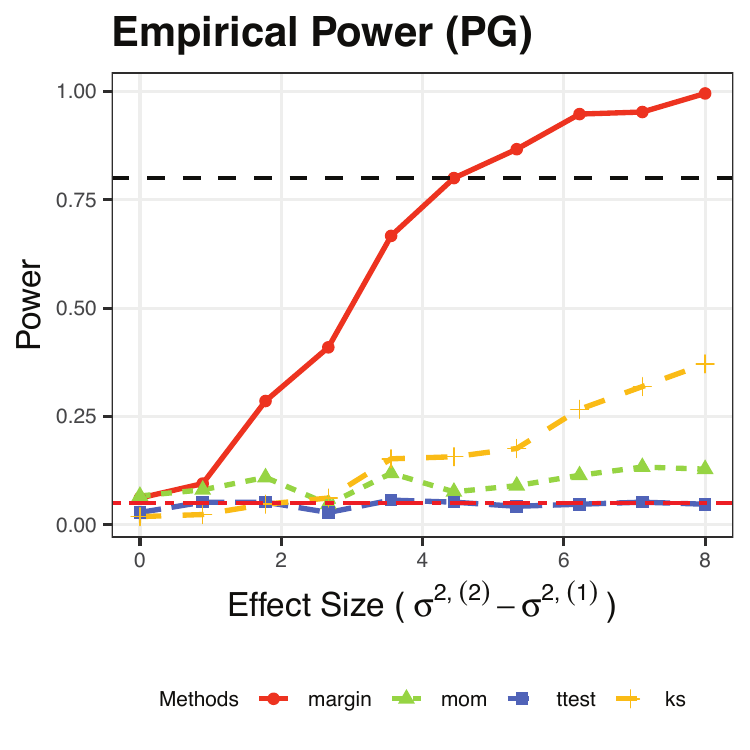} 
	\includegraphics[width = 0.45\textwidth,height=0.3\textheight]{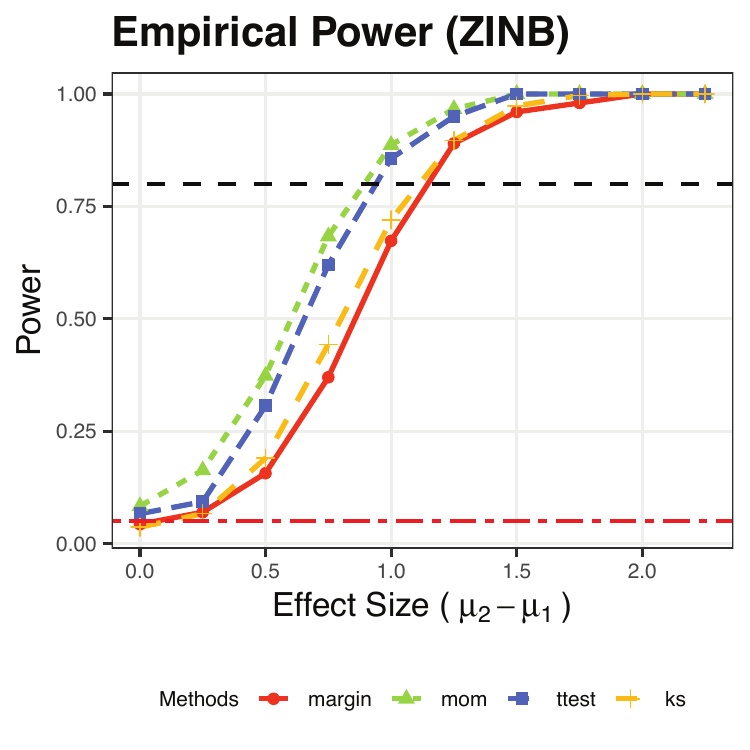}
	\includegraphics[width = 0.45\textwidth,height=0.3\textheight]{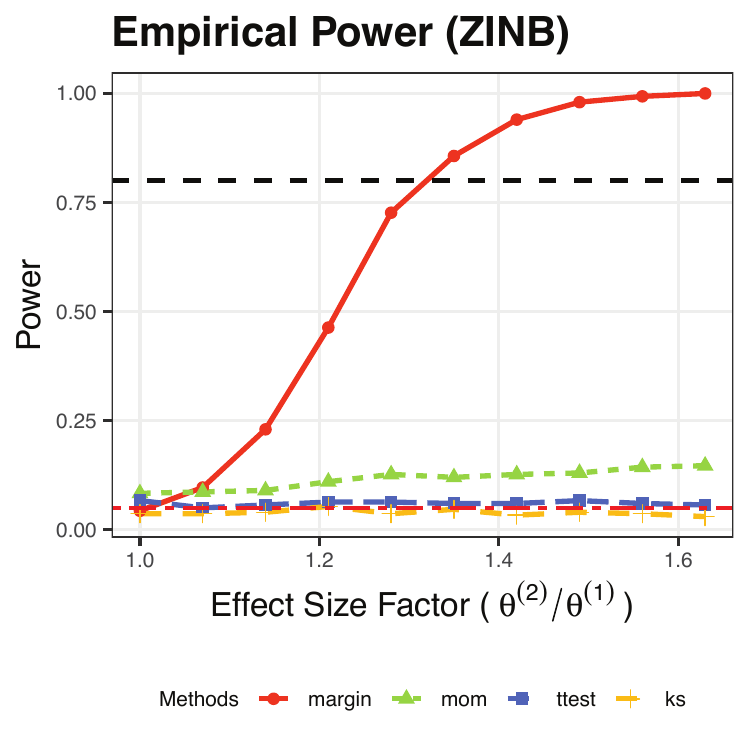} 
	\caption{Simulation results. Top panel: QQ-Plot of negative log-transformed p-values for Poisson-Gamma model (left), and ZINB (right) when distributions are equivalent. Middle panel: power comparison for Poisson-Gamma model under mean and variance shift. Bottom panel: power comparison for ZINB model under mean/variance shift  and variance shift.
		Results use 300 simulation experiments with sample size $n_1 = n_2 = 100$.}
	\label{fig:simQQplot}
\end{figure}

For both models, we make a few observations. First, under mean shift settings, the proposed SEF test exhibits competitive performance with a pseudobulk t-test or K-S test (see Figure~\ref{fig:simQQplot}, middle and bottom panels). Changes in the mean are easier to detect with pseudobulk approaches, so we expect similar performance under this model setting. 

Second, under variance shift settings, it is clear that our proposed SEF test exceeds performance of these competing methods for all models. Our power analysis under Poisson-Gamma or zero-inflated settings indicates that pseudobulk tests fail to demonstrate any sort of ability to detect changes in variance, and the standard moment estimator unable to render any performance  (Figure \ref{fig:simQQplot}, middle and bottom panels). Our results under the Poisson-Gamma and zero-inflated models indicate the SEF test holds a clear performance advantage over existing methods in detection of distributional changes beyond the first moment. Just as importantly, the SEF test's performance is indiscriminate of the underlying distribution of the data, ranging from the Poisson-Gamma cases to overdispersed count models.
It should be noted that under sparser scenarios (such as the zero-inflated setting), the standard moment estimator struggles with performance significantly in comparison to the SEF test (see \ref{supp:MoMvsSEF}).


For additional simulation settings, including unequal sample sizes (Figure \ref{FigS_n1n2}), difference in higher moments (Figure \ref{FigS_p3}) and comparisons with pseudo-bulk F-test of variances (Figure \ref{FigS_F}), see \ref{supp:numericalExp} for details. 

From these simulation results, we conclude the SEF test operates at a higher level than existing classic methods under higher moment settings across various data types due to its model-agnostic nature, and does so all within a single unified framework.

\section{Application to SLE Single-Cell RNA-seq Data}
We apply the proposed SEF framework to  a population scale single cell RNA-sequencing data set from~\cite{perez2022}, profiling peripheral blood mononuclear cells (PBMCs) from 256 samples (156 SLE cases 98 healthy controls and 483,336 cells). The original study contains over 1.2 million cells (759,202 SLE and 483,336 healthy control cells), spanning 11 annotated cell types; we focus on three abundant cell types: classical monocytes, CD4+ and CD8+ T-cells, owing to their abundance among the dataset (see Table~\ref{tab:SLE-prop}). These cell types are biologically of interest due to  the roles they play in autoimmune disease. Monocytes are active in many cellular processes, including cytokine production and apoptosis of cells \citep{kavai2007immune, henriques2012functional}. CD4+ and CD8+ T-cells are also contributors to immune response in SLE, having been shown to be directly associated to antibody production, cytoxicity to target cells, and cell-to-cell contact-mediated suppressive responses \citep{suzuki2008inhibitory, suarez2016, li2022cd8}.
Our objective is to identify genes with significant, cell type-specific distributional differences between SLE and control samples.

\subsection{Preprocessing}
We apply standard quality control at both the gene and cell levels. Using cell-type annotations from \cite{perez2022}, we normalize and batch-correct these data using \texttt{SCTransform} integration within the Seurat pipeline in \texttt{R} \citep{stuart2019seurat, choudhary2022sctransform}. Donors were excluded if, for any gene, they had fewer than 100 cells or less than 20\% nonzero counts. From this, genes with fewer than 50 donors in either disease group were discarded. Library size-adjusted counts were used as a measure of gene expression in our analysis. 
While many existing methods employ a log-transformation with a pseudocount (i.e $\log (y + 1)$), it has become evident that this introduces distortions to the underlying expression distribution \citep{townes2020}.
As such, we employ a square-root plus $1/2$ (i.e., $\sqrt{y+1/2}$) transformation to the corrected counts to account for heavy-tails \citep{bartlett1936}.

\subsection{Comparison of Genes Identified With Different Methods}
After preprocessing, we have 907, 596, and 605 genes in total for classical monocytes, CD4+ T-cells, and CD8+ T-cells, respectively. Of these genes, 299, 239, and 249 exhibited statistically significant distributional differences for each cell type, in the same order. Example genes highlighted in Figure~\ref{fig:grpComparison} are based on p-value and gene ontology (GO) canonical pathways related to immune and inflammatory processes and responses for each cell type. We note in particular that many of these genes, namely ISG15, HLA-A, and B2M, are reliable markers for SLE in studies \citep{perez2022, aghdashi2019B2M}. We also include S100A9 in monocytes, NKG7 in CD8+ T-cells, and both LTB and IL32 in CD4+ T-cells. S100A9 is included in Figure~\ref{fig:grpComparison} due to longstanding literature in rheumatology supporting the association between exacerbated expression levels in patients and SLE within the S100 family, with S100A8/9 in particular \citep{soyfoo2009s100a9}. This sentiment is also evident in IL32, as CD4+ T-cells often experience enhanced expression of the gene in autoimmune contexts \citep{de2021IL32}. NKG7 is incorporated into the figure due to its well-defined roles in apoptosis and inflammation in autoimmune and cancer contexts, and is also discussed in the original study as a cytotoxic marker in CD8+ T-cells \citep{ng2020nkg7, perez2022, wang2025nkg7Cytotoxic}. LTB plays a role in adaptive immune response for T-cells and can support T-cell differentiation~\citep{upadhyay2013LTBdifferentiationTcell}.

For each cell type of interest, we demonstrate the differing gene expression signatures between SLE and healthy samples in canonical enrichment pathways. These distributional differences among these genes are illustrated in Figure \ref{fig:grpComparison}. Importantly, group-wise density comparisons reveal clear variability and location shifts in expression between the two populations. The phenomena of autoimmune disease or other perturbations, such as aging, affecting the variability of gene expression is well-established in other single-cell studies~\citep{enge2017agingvariation}, and sheds light onto how SLE lupus affects expression at the distribution level. Our results suggest that the SEF test can identify biologically meaningful genes relevant to the immune responses to SLE lupus in humans. See Supplementary Materials~\ref{sec:supp_RDA} for additional visualization on carrier density estimate accuracy (Figure \ref{supp:carrierdens}) and individual-specific model fitting of these genes (Figure \ref{supp:indiv_density}). 

As a means of comparison, we look for differentially expressed genes using MAST and a standard pseudobulk Wilcoxon rank-sum test, two of the most common methods in existing scDE analysis pipelines. We remark that while DESeq2 and edgeR are also frequently used methods in the pseudobulk context, the population-scale nature of our study lends itself to better inference using a Wilcoxon rank-sum test \citep{li2022wilcoxon_DE}.
MAST leads to large inflation with  more than 80\% of the total genes tested across each cell type being  differentially expressed (767, 504, 524 genes for monocytes, CD4+, and CD8+ T-cells respectively). As a result of its inability to control for false positives, our downstream comparisons omit MAST.

\begin{table}[h!]
	\caption{Comparison of DEGs between SEF test and pseudobulk approach for all three cell types. Diagonal entries represent the total number of uniquely identified genes, and off-diagonals represent genes found by both methods. }
	\label{fig:SEF-PB-degTables}
	\centering
	\begin{subfigure}[b]{0.32\textwidth}
		\centering
		\caption*{Monocytes}
		\begin{tabular}{c|cc}
			& SEF & PB \\
			\hline
			SEF & 61 & 238 \\
			PB  & 238 & 42 \\
		\end{tabular}
	\end{subfigure}
	\begin{subfigure}[b]{0.32\textwidth}
		\centering
		\caption*{CD8+}
		\begin{tabular}{c|cc}
			& SEF & PB \\
			\hline
			SEF & 79 & 170 \\
			PB  & 170 &  32 \\
		\end{tabular}
	\end{subfigure}
	\begin{subfigure}[b]{0.32\textwidth}
		\centering
		\caption*{CD4+}
		\begin{tabular}{c|cc}
			& SEF & PB \\
			\hline
			SEF & 37 & 202 \\
			PB  & 202 &  42 \\
		\end{tabular}
	\end{subfigure}
\end{table}

We use an aggregated pseudobulk Wilcoxon rank-sum test as another method to compare with the SEF test, yielding 280, 244, and 202 genes for monocytes, CD4+, and CD8+ T-cells, respectively. Though total number of genes  identified  are comparable  between the SEF and pseudobulk Wilcoxon and the genes identified largely overlap, each method also identifies its own set of genes (see Table \ref{fig:SEF-PB-degTables}). However, the genes only identified by the pseudobulk Wilcoxon test show very small fold changes ($<0.1$) (see Figure \ref{fig:fold}).

\begin{figure}
	\centering
	\includegraphics[width=0.32\textwidth]{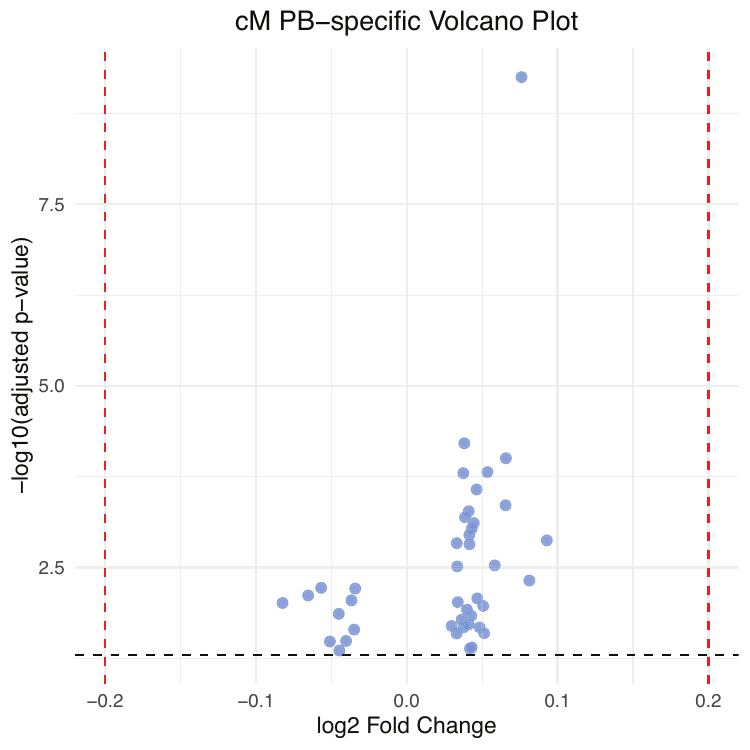}	
	\includegraphics[width=0.32\textwidth]{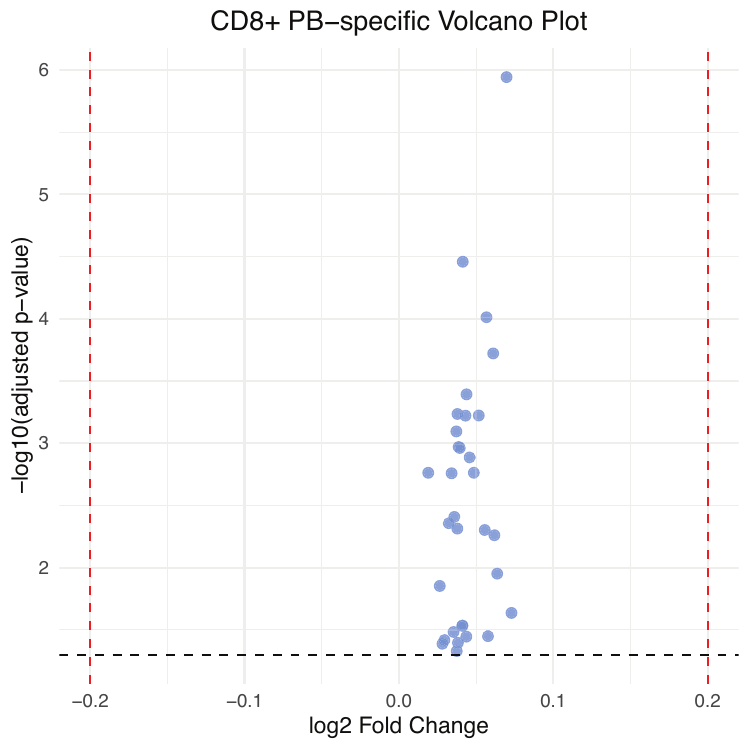}
	\includegraphics[width=0.32\textwidth]{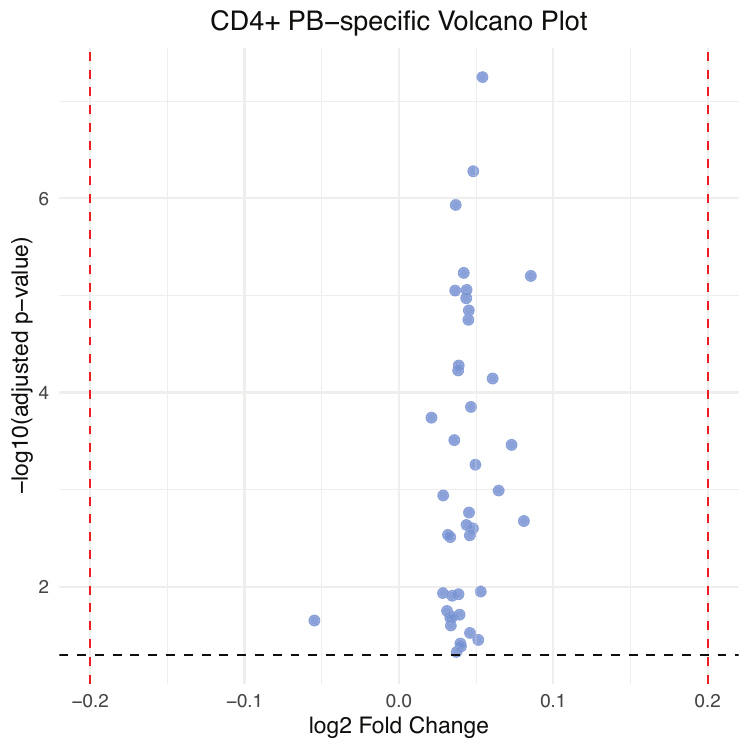}
	\caption{Volcano plots of genes only identified by pseudobulk Wilcoxon test  for classical monocytes (left), and CD8+ T-cells (middle), and CD4+ T-cells (right).}
	\label{fig:fold}
\end{figure}

We further verify that SEF is able to uniquely identify genes with more meaningful biological signatures. 
For CD4+ T-cells, we experience the same phenomena, with the proposed SEF framework yielding 37 uniquely identified genes, including TSC22D3, HLA-DRA, FOS, and RGS10. These genes provide a variety of immune function, with RGS10 directly related to T-cells and adaptive immune response, TSC22D3 (or the directly associated protein GILZ) being a contributor to modulating T-cell apoptosis, and the MHC class II gene HLA-DRA playing a crucial part specifically in CD4+ T-cells in presenting antigens \citep{thul2017HumanProteinAtlas, cannarile2019TSC22D3_GILZ}.
Furthermore, the Human Protein Atlas suggests that FOS contributes to cell differentiation, proliferation and transduction signaling, and recent studies suggest that upregulation in FOS signatures is a consequence of CD4+ T-cell activation due to infection \citep{jimenez2023FOS,chang2025cd4FOS}. 
The SEF framework identifies 79 unique CD8+ T-cell-specific genes of interest, many of which contribute to lymphocyte and cytotoxic activity--- processes in autoimmune diseases that induce aberrant killing of cells. We highlight the unique identification of S100A4, IL7R, LTB, and MT2A as interesting genes demonstrating potential chronic, long-term responses to SLE \citep{wong2025cd8_s100a4}. It has been proven that increases in lymphotoxin-$\beta$ and serum-soluble interleukin 7, proteins encoded by LTB and IL7R, are significant markers of SLE due to their roles in the adaptive immune response \citep{badot2013IL7R, yin2013LTB}. The SEF test uniquely identifies and characterizes the differences in the distributional expression of these two protein-coding genes. Despite suppressed expression of LTB in CD8+ T-cells, the subtle difference between the tails of expression densities potentially suggests T-cell exhaustion over time, which has been seen in cases of chronic inflammation \citep{yi2010tcellExhaustion,lima2021exhaustedSLE}. In particular, LTB demonstrates more variable non-zero expression in SLE patients than healthy controls within CD4+ helper T-cells (Figure \ref{fig:grpComparison}), suggesting that the higher zero-mass in gene expression density among the CD8+ T-cell population may have induced exhaustion. MT2A, meanwhile, has long been viewed as a gene related to reducing apoptosis and oxidative stress \citep{ling2016MT2A}. The lower zero-inflated mass for the SLE group in the expression curve of MT2A suggests a potential reaction to the significant stress and cell-killing induced by SLE. 
Lastly, within the classical monocytes cell type, we see that SEF contains 61 uniquely identified genes, including important monocyte-specific innate immune response-related genes such as S100A4, S100A8, RGS2, and MAFB, as seen in the Human Protein Atlas \citep{thul2017HumanProteinAtlas}. In particular, higher and more variable levels of S100A4 in monocytes and the oligomer oS100A4 have associations to other autoimmune related diseases, such as rheumatoid arthritis (RA) \citep{neidhart2019oS100A4}. Furthermore in the S100 family, not only has S100A8 been seen to exhibit raised expression levels in SLE patients as mentioned earlier, but the calcium-binding protein-coding gene can also be differentially expressed in lupus nephritis, a major comorbidity of SLE~\citep{soyfoo2009s100a9, qijiao2022s100a8LupusNephritis}.
In the Supplementary materials, we provide visualizations of the group comparisons for the aforementioned uniquely-identified genes of interest (see S100A4/8, TSC22D3, MT2A, and LTB in Supplementary Figure \ref{supp:unique-DEG-viz}). 

\begin{figure}
	\centering
	\includegraphics[width=0.32\textwidth]{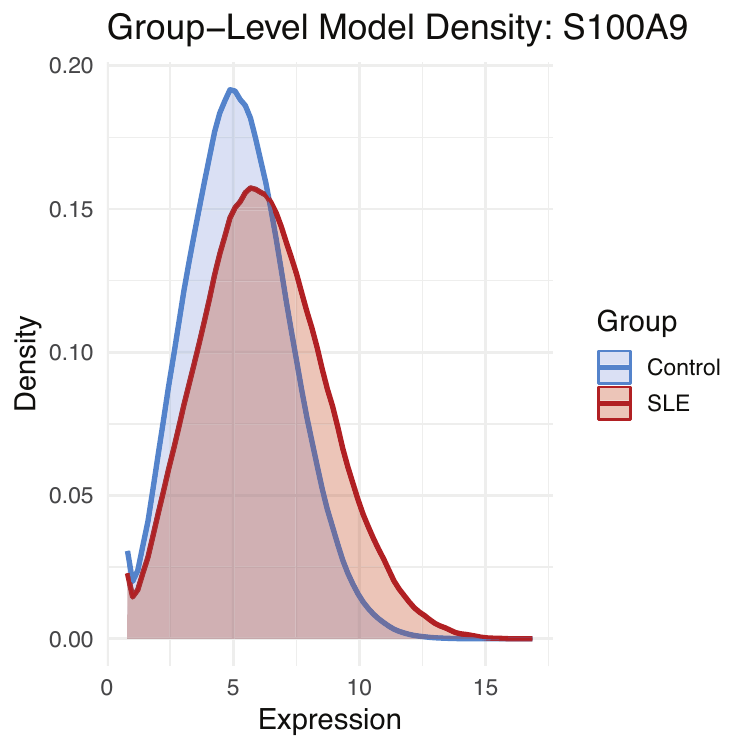} 
	\includegraphics[width=0.32\textwidth]{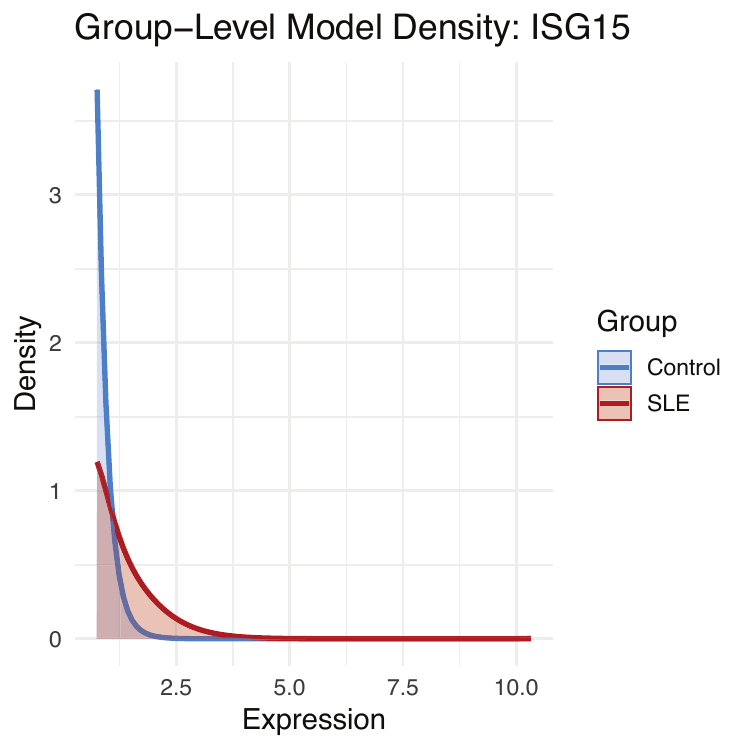}
	\includegraphics[width=0.32\textwidth]{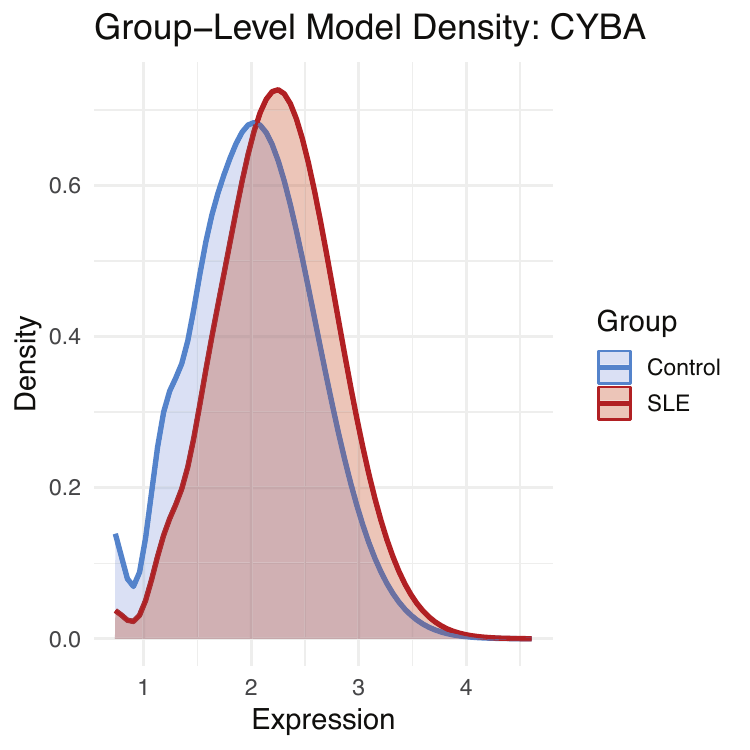}   \\
	
	(a) Monocytes
	\vspace{0.5em}
	
	\includegraphics[width=0.32\textwidth]{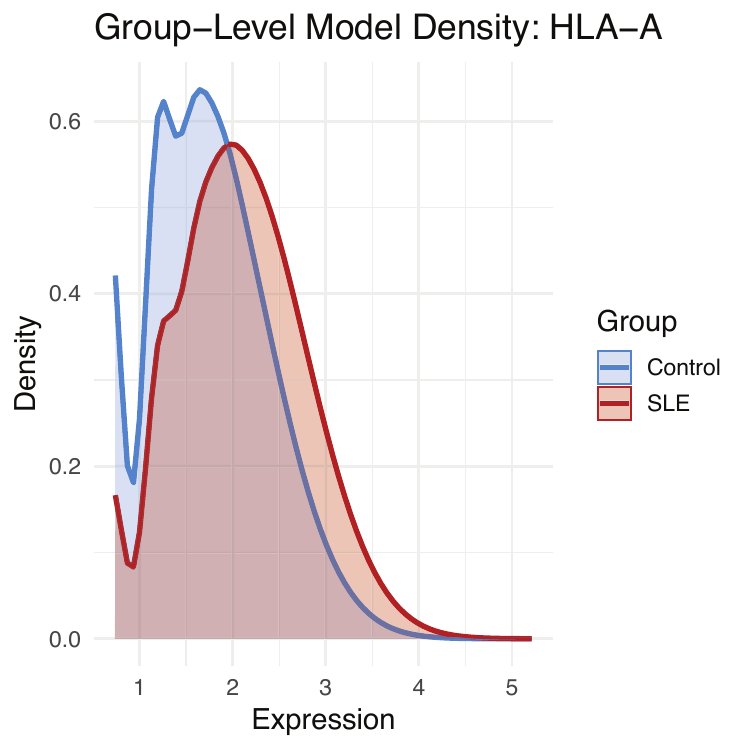} 
	\includegraphics[width=0.32\textwidth]{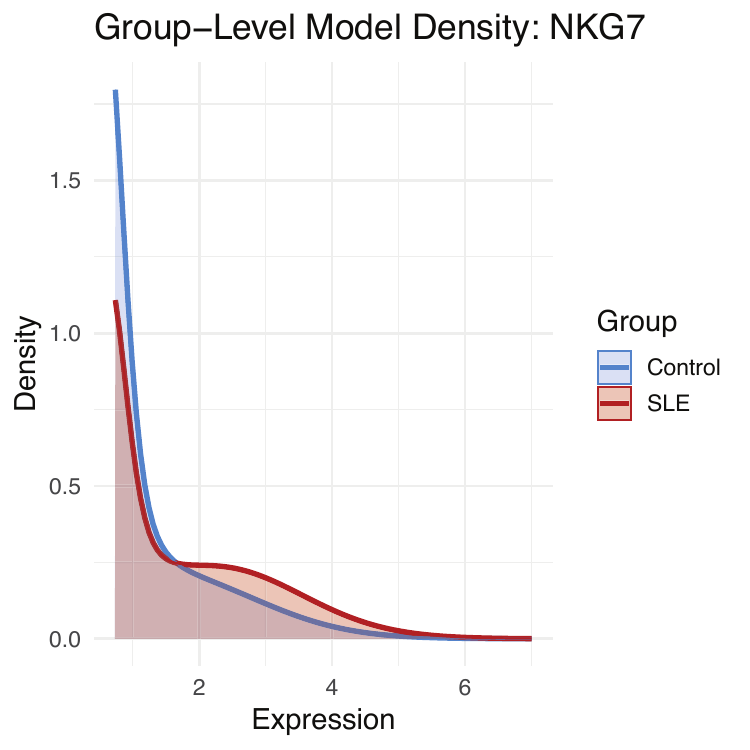} 
	\includegraphics[width=0.32\textwidth]{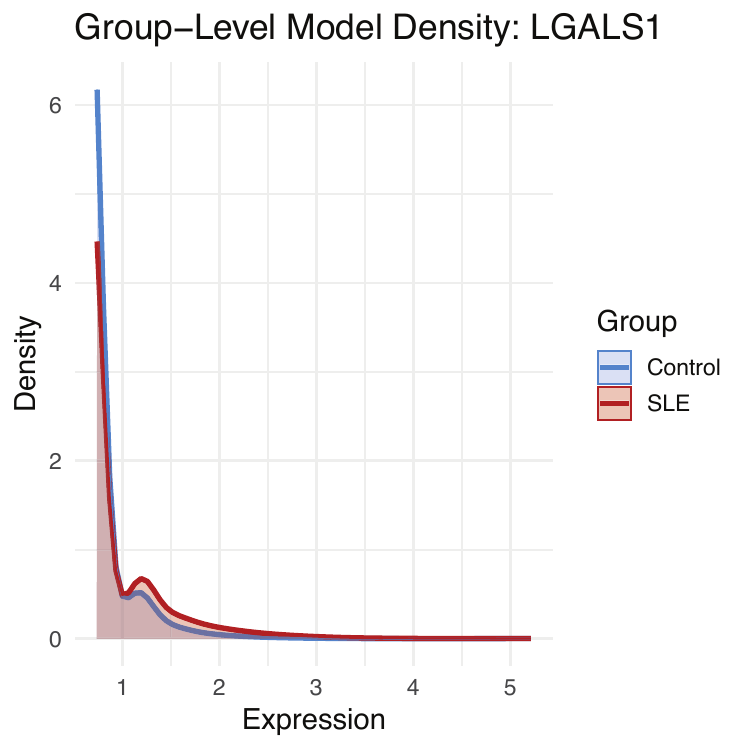}   \\
	
	(b) CD8+ T-cells
	\vspace{0.5em}
	
	\includegraphics[width=0.32\textwidth]{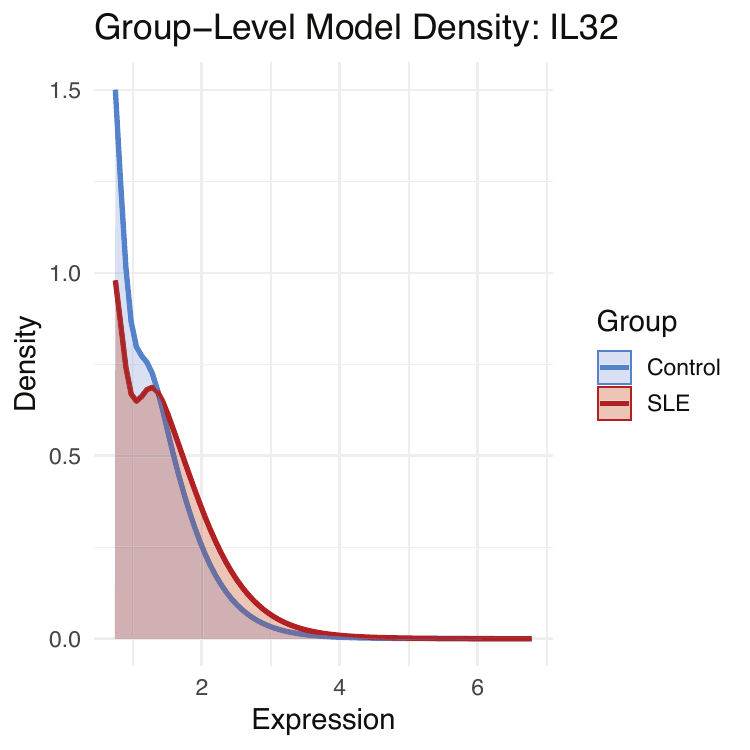} 
	\includegraphics[width=0.32\textwidth]{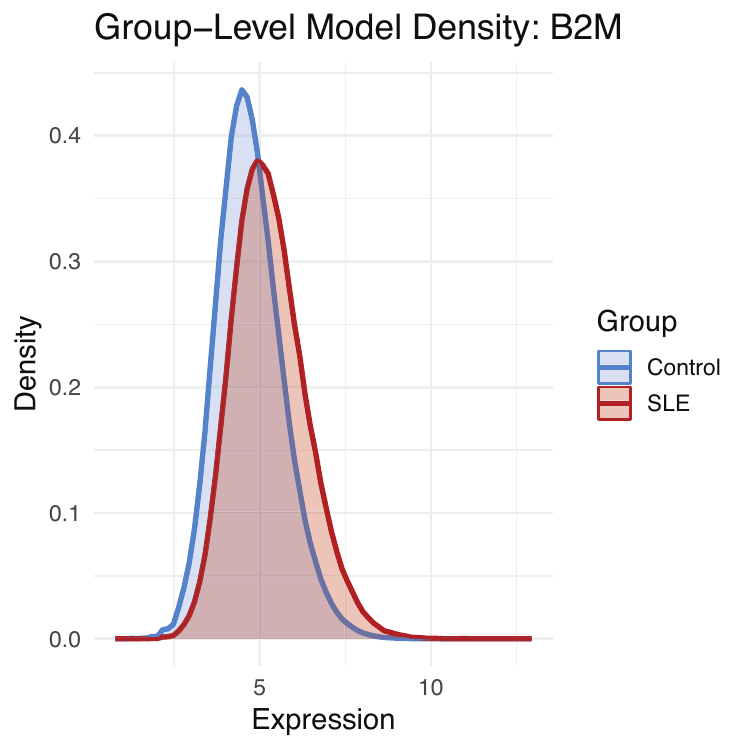} 
	\includegraphics[width=0.32\textwidth]{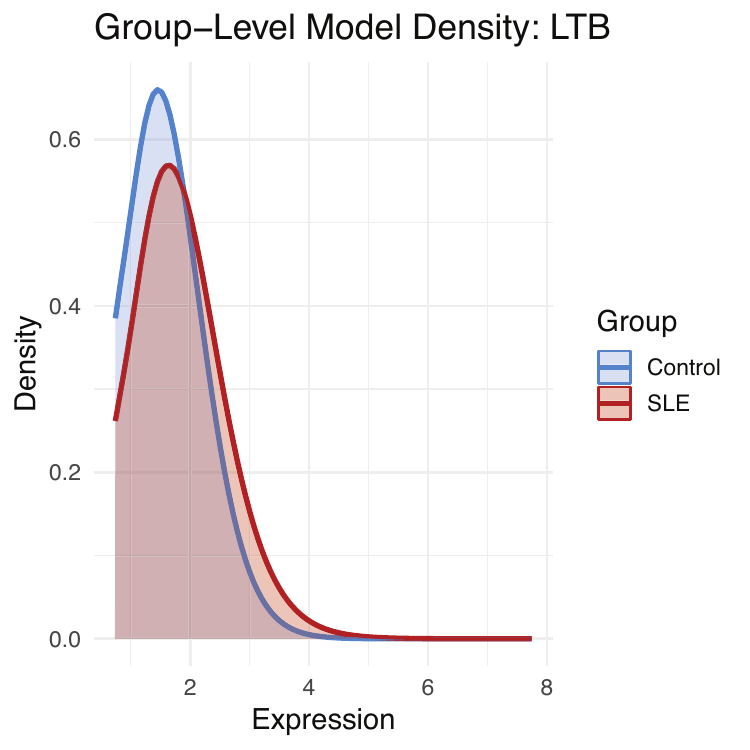}\\
	
	(c) CD4+ T-cells
	\vspace{0.5em}  
	
	\caption{Comparison of estimated group-wise expression densities between SLE (red) and healthy (blue) groups. Genes shown are from biologically relevant canonical pathways for classical monocytes (row 1), CD8+ T-cells (row 2), and CD4+ T-cells (row 3). We also include genes previously shown to have clear associations to SLE (S100A9, NKG7, IL32, LTB) throughout immunology and rheumatology literature. All genes shown demonstrate adjusted p-values $\approx 0$, with the exception of B2M and LGALS1 with p-values at $7.940\times10^{-13}$ and $4.702\times 10^{-13}$, respectively.}
	\label{fig:grpComparison}
\end{figure}

We provide a final means of comparison between pseudobulk Wilcoxon and the SEF test by computing the proportion of DEGs found by each method that are also validated from the supplementary DE analysis in the original \cite{perez2022} study. Using the intersection of genes tested for each cell type and a list of cell type-specific genes found in the original study, it becomes evident that the SEF test identifies a higher proportion of these literature-validated genes than the pseudobulk Wilcoxon (See Table \ref{fig:Perez-validation-table}).

\begin{table}[h!]
	\caption{Proportion of DEGs found using the SEF test and pseudobulk Wilcoxon that are validated by the original \cite{perez2022} supplementary study for three cell types.}
	\label{fig:Perez-validation-table}
	\centering
	\begin{tabular}{c|ccc}
		& Monocytes & CD8+ & CD4+ \\
		\hline
		SEF & 0.535 & 0.680 & 0.739 \\
		PB (Wilcoxon) & 0.441 & 0.520 & 0.565 \\
	\end{tabular}
	
\end{table}

To assess false discovery control, we performed permutation experiments by randomly shuffling disease labels and re-running the SEF analysis. Then, we conduct hypothesis testing for the set of genes of each cell type. QQ-plots of the resulting p-values in Figure~\ref{fig:semisimu} for 10 permutation experiments were close to the expected uniform distribution, and the proportion of false positives was near the nominal 5\% in all cell types. 
\begin{figure}
	\centering
	\includegraphics[width=0.32\textwidth]{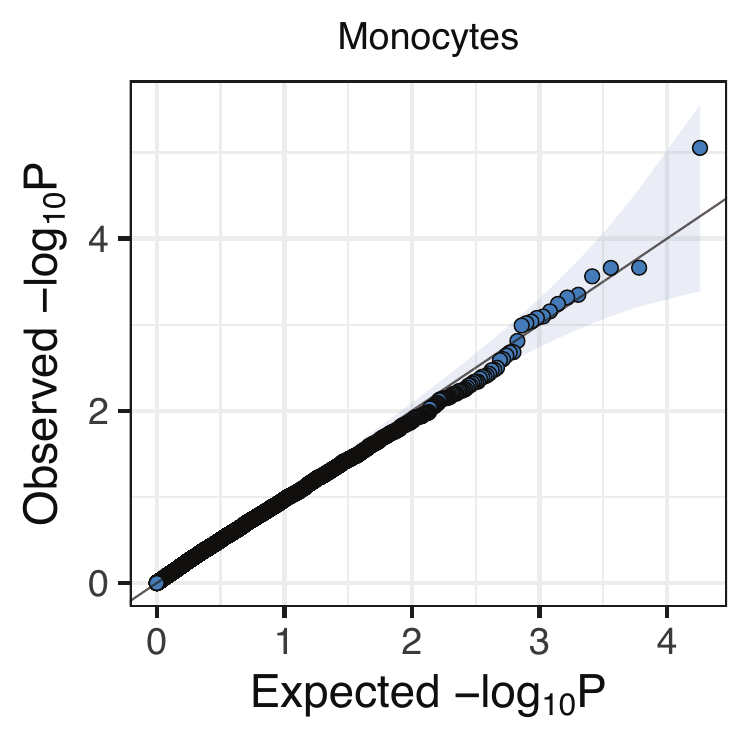}
	\includegraphics[width=0.32\textwidth]{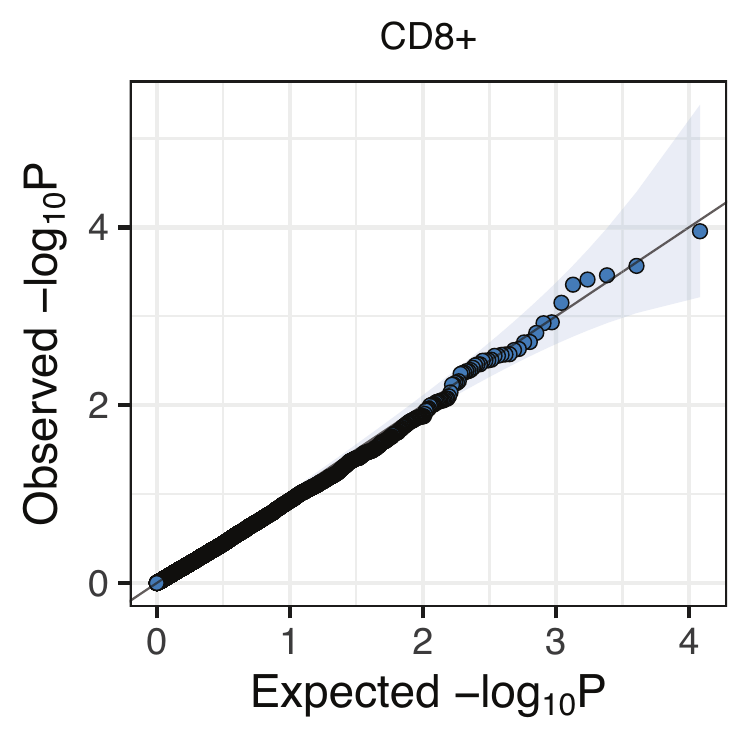}
	\includegraphics[width=0.32\textwidth]{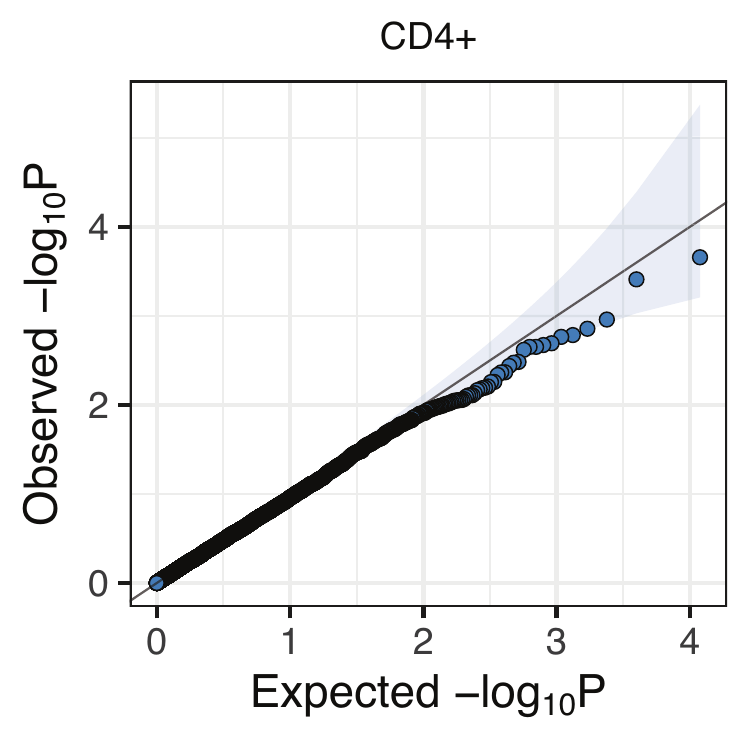}
	\caption{QQ-Plot of negative log-transformed p-values in disease label-permuted test among cells from classical monocytes (left), and CD8+ T-cells (middle), and CD4+ T-cells (right). A total of 10 numerical experiments were run, with figures displaying p-values for each gene tested (9070, 6050, and 5960 genes, respectively).}
	\label{fig:semisimu}
\end{figure}

\subsection{Gene Set Enrichment Analysis}
A natural follow-up in single cell downstream analysis after DEG identification is gene enrichment analysis. In particular, we use a gene set over-representation analysis (ORA) in the \texttt{clusterProfiler} package to demonstrate that the SEF framework captures holistic and biologically-meaningful signatures (Figure~\ref{fig:enrichment}). Genes found using the SEF framework highlight classic immune regulatory and effector signatures across cell types of interest, such as proliferation, differentiation, and regulation of T-cells, leukocyte, and lymphocytes. Our enrichment results confirm the biological meaningfulness of the cell type-specific genes of interest found by the SEF test, seeing that monocyte pathways correctly highlight innate immune responses and leukocyte proliferation along with CD4+ pathways confirming regulatory and activation of adaptive response cells such as regulatory T-cells or lymphocytes.
Furthermore, SEF-derived enrichment analysis identifies pathways in monocytes that are related to T-cell interactions, with antigen processing and presentation in particular. The relationship between monocytes and T-cells regarding autoimmune disease has since been established in literature~\citep{tu2021monocyte_tcell_crosstalk}. The SEF-derived pathways of antigen processing and presentation along with endocytosis regulation supports the suggestion that classical monocytes can promote T-cell proliferation and activation, mirroring these findings. Lastly, enrichment results from within the CD8+ T-cell type further reinforce the potential phenomena of T-cell exhaustion. In particular, regulatory signatures of lymphocyte apoptosis and T-cell activation reinforce this idea, as such exhaustion can occur from dysfunction due to chronic stimulation. Further, leukocyte-mediated cytotoxicity and positive leukocyte cell-cell adhesion can exacerbate the dysfunctionally persistent immune response.


\begin{figure}
	\centering
	\includegraphics[width=0.32\textwidth]{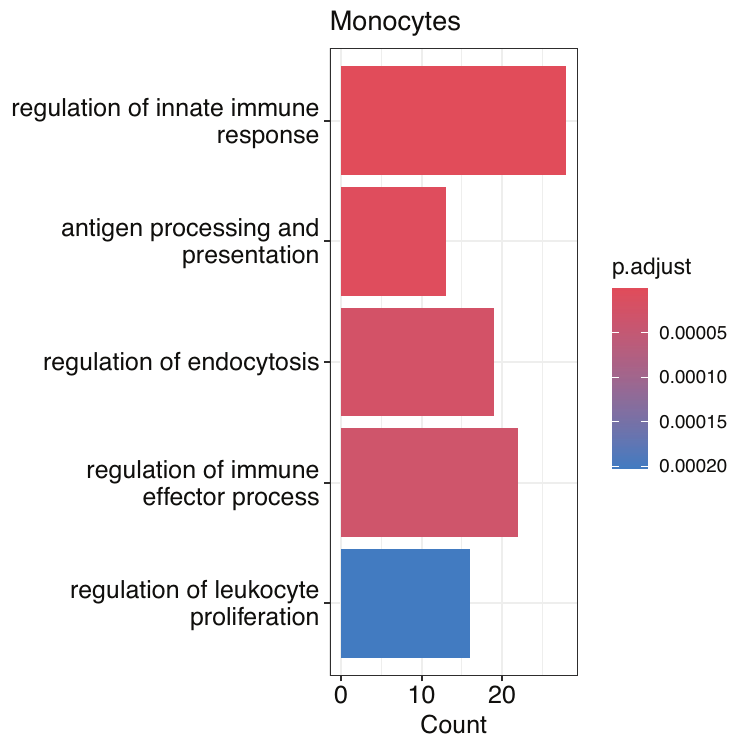}
	\includegraphics[width=0.32\textwidth]{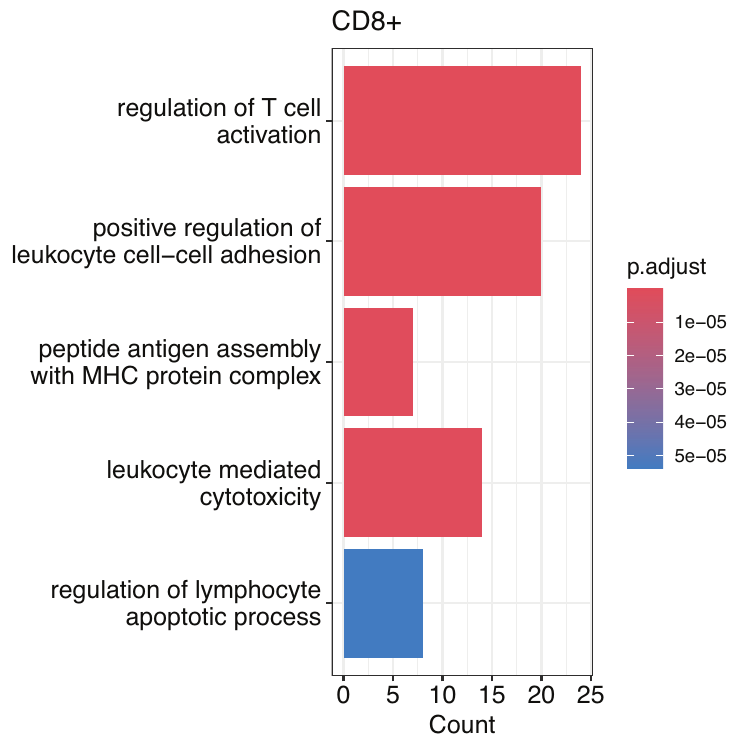}
	\includegraphics[width=0.32\textwidth]{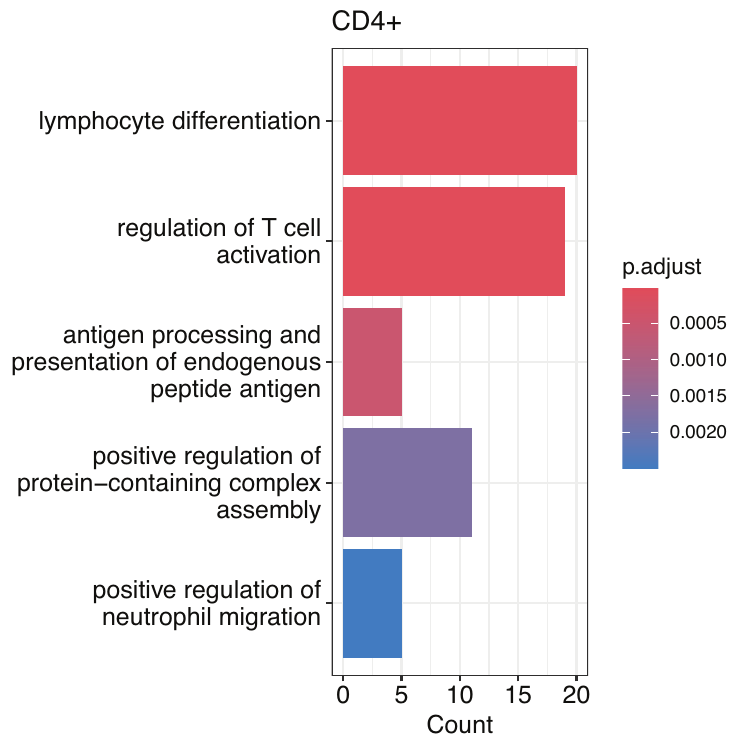}
	\caption{GO Enrichment Pathways of Interest for each cell type (starting left to right: monocytes, CD8+ T-cells, CD4+ T-cells). Processes shown are from top 20 most significant pathways using \texttt{clusterProfiler} software package. P-values were adjusted using Benjamini-Hochberg correction with a threshold of 0.05.}
	\label{fig:enrichment}
\end{figure}

\section{Conclusion}
In this work, we introduced a specially designed exponential family (SEF) framework for population-scale single-cell analysis that models each individual’s gene expression as a probability density. By moving beyond mean expression and accommodating diverse underlying data-generating mechanisms, SEF provides a flexible, model-agnostic approach for density estimation, visualization, and formal inference. Our theoretical analysis established the validity of the method through asymptotic guarantees and a consistent covariance estimator, while simulations demonstrated improved power and error control compared to existing approaches such as pseudo-bulk tests, moment-based estimators and MAST. 

The application to systemic lupus erythematosus (SLE) scRNA-seq data illustrates the biological relevance of SEF, uncovering genes and expression patterns that may not be captured by conventional mean-based methods. This highlights the potential of density-based approaches to reveal subtle, yet biologically meaningful, differences in expression distributions across individuals and groups, particularly in the context of complex diseases where cell-type heterogeneity and expression variability in cells of the same type are key.

Several avenues for future work emerge from this study. First, while SEF provides a general and robust framework, extensions to incorporate covariates, longitudinal designs, and multi-omic single-cell data would further broaden its utility. Second, exploring computational strategies for large-scale data integration and distributed inference will be critical as population-scale scRNA-seq studies continue to expand. Finally, integrating SEF-derived density estimates with downstream tasks such as clustering, trajectory inference, or causal modeling may open new opportunities for discovery in precision medicine.

In summary, SEF offers a principled and practical framework for population-level density-based inference in scRNA-seq studies. By enabling rigorous hypothesis testing and biologically interpretable density comparisons, our approach provides a foundation for more comprehensive understanding of cellular heterogeneity in health and disease.


\section*{FUNDING}
This research was supported by NIH grants U01HG013841 and R01GM129781. 

\section*{SUPPLEMENTARY MATERIALS}

In the Supplementary Materials, we prove all the main theorems and the technical lemmas. We also provide additional simulation results and details of real data analysis.

\bibliographystyle{chicago}
\bibliography{Reference}

\end{document}